\pdfoutput=1

\documentclass[twocolumn,final]{svjour3}

\smartqed

\usepackage{graphicx,amssymb}
\usepackage{cases}
\usepackage{epsf,psfig}
\usepackage{txfonts}
\usepackage{multirow}
\usepackage{bigstrut}

\journalname{Nonlinear Dynamics}

\begin{document}

\title{The Fast Norm Vector Indicator (FNVI) method: A new dynamical parameter for
detecting order and chaos in Hamiltonian systems}

\author{Euaggelos E. Zotos}

\institute{Euaggelos E. Zotos:
\at Department of Physics, \\
Section of Astrophysics, Astronomy and Mechanics, \\
Aristotle University of Thessaloniki \\
GR-541 24, Thessaloniki, Greece \\
email:{evzotos@physics.auth.gr}
}

\date{Received: 8 May 2012 / Accepted: 7 June 2012 / Published online: 30 June 2012}

\titlerunning{Detecting order and chaos in Hamiltonian systems by the Fast Norm Vector Indicator (FNVI) method}

\authorrunning{Euaggelos E. Zotos}

\maketitle

\begin{abstract}

In the present article, we introduce and also deploy a new, simple, very fast and efficient method, the Fast Norm Vector Indicator (FNVI) in order to distinguish rapidly and with certainty between ordered and chaotic motion in Hamiltonian systems. This distinction is based on the different behavior of the FNVI for the two cases: the indicator after a very short transient period of fluctuation displays a nearly constant value for regular orbits, while it continues to fluctuate significantly for chaotic orbits. In order to quantify the results obtained by the FNVI method, we establish the dFNVI, which is the quantified numerical version of the FNVI. A thorough study of the method's ability to achieve an early and clear detection of an orbit's behavior is presented both in two and three degrees of freedom (2D and 3D) Hamiltonians. Exploiting the advantages of the dFNVI method, we demonstrate how one can rapidly identify even tiny regions of order or chaos in the phase space of Hamiltonian systems. The new method can also be applied in order to follow the time evolution of sticky orbits. A detailed comparison between the new FNVI method and some other well-known dynamical methods of chaos detection reveals the great efficiency and the leading role of this new dynamical indicator.


\end{abstract}

\section{Introduction}

The issue of knowing whether the orbits of a dynamical system are ordered or chaotic is fundamental for the understanding of the behavior of the system in an extended area of modern science. In the dissipative case, this distinction is easily made as both types of motion are attracting. On the other hand, in conservative systems, distinguishing between regular and chaotic motion is often a very delicate and difficult issue (i.e. when the chaotic or ordered regions are small), especially in dynamical systems with many degrees of freedom, where one cannot easily visualize and interpret directly the behavior of the orbits. For this reason, it is often of great importance to possess fast and accurate tools in order to determine if an orbit is ordered or chaotic, independent of the dimension of the phase space of the dynamical system.

Let's recall and present some well-known, classical methods that try to give an answer to the issue of determining the nature of an orbit.

\textbf{(a)} The inspection of the consequents of an orbit on a Poincar\'{e} surface of section (PSS). For 2D dynamical systems this technique has been used extensively, despite the problem of establishing a proper Poincar\'{e} surface of section in each case. However, the inspection becomes very difficult and also greatly deceiving in the case of dynamical systems with multidimensional phase space.

\textbf{(b)} The maximal Lyapunov Characteristic Exponent (LCE) $\sigma$ of an orbit informs us whether an orbit is ordered or chaotic. If $\sigma > 0$ then the corresponding orbit is chaotic. Over thirty years ago, Benettin et al [2] studied theoretically the problem of the computation of all LCEs and proposed an algorithm for their numerical computation. In particular, $\sigma$ is computed as the limit for $t \rightarrow \infty$ of the quantity
\begin{equation}
L_t = \frac{1}{t} \ln \frac{\|\vec{w}(t)\|}{\|\vec{w}(0)\|},
\end{equation}
where $\vec{w}(0)$ and $\vec{w}(t)$ are the deviation vectors for a given orbit, at times $t = 0$ and $t > 0$ respectively. The time evolution of $\vec{w}$ is given by solving the so-called \textit{variational equations}. Generally, for almost all choices of initial deviations $\vec{w}(0)$, the limit for $t \rightarrow \infty$ of equation (1) gives always the same value of $\sigma$
\begin{equation}
\sigma = \lim_{t\to\infty} L_t.
\end{equation}

In practice, of course, since the exponential growth of $\vec{w}(t)$ occurs for short time intervals, one stops the evolution of $\vec{w}(t)$ after some time $T_1$, records the computed $L_{T_1}$, normalize the vector $\vec{w}(t)$ and repeats the calculation for another time interval $T_2$, etc obtaining finally the $\sigma$ as an average over many $T_i$, $i = 1, 2, . . . , N$ as
\begin{equation}
\sigma = \frac{1}{N} \sum_{i=1}^{n} L_{T_i}.
\end{equation}
The basic drawback of the computation of $\sigma$ is that, after every $T_i$, the calculation starts from the beginning and may yield an altogether different $L_{T_i}$ than the $T_{(i-1)}$ interval. Thus, since $\sigma$ is influenced by the whole evolution of $\vec{w}(0)$, the time needed for $L_t$ (or the $L_{T_i}$) to converge is not known \textit{a priori} and may become extremely long. This makes it often very difficult to determine whether $\sigma$ finally tends to a positive value (chaos) or converges to zero (order). The main advantage of the LCE is that it can be applied easily to dynamical systems of any number of degrees of freedom.

\textbf{(c)} The frequency analysis method proposed by Laskar [14-16, 19, 20], which is based on the calculation of the basic frequencies of an orbit over a fixed interval of time. For orbits on the KAM tori, these frequencies are very accurate approximations of the actual frequencies of the dynamical system, but for chaotic orbits the computed values vary significantly in time and space. The frequency analysis method can be applied to systems with many degrees of freedom.

\textbf{(d)} The study of the dynamical spectra of orbits [6, 21, 28, 34, 35]. The distribution of the values of a given parameter along the orbit has been proved a very reliable and powerful tool for the exploration of the properties of motion in Hamiltonian systems of two and three degrees of freedom. The reader can find more interesting information about the dynamical spectra in [35] and also in the references of this article.

Over the last years, several methods have been introduced in order to characterize the nature of an orbit by studying the evolution of the deviation vectors, some of which are discussed in section 4. In the present paper we introduce and use a new, fast and easy to compute indicator: the Fast Norm Vector Indicator (FNVI). We focus our attention on the method of the FNVI, performing a systematic and thorough study of its behavior in the case of autonomous Hamiltonian systems with two (2D) and three (3D) degrees of freedom. The FNVI was found to fluctuate significantly for chaotic orbits, while it displays a nearly constant value for ordered orbits.

This article is organized as follows: in section 2 we provide the definition of the FNVI and we present results distinguishing between ordered and chaotic motion in two and three degrees of freedom (2D and 3D) Hamiltonians, comparing also the efficiency of the FNVI with the computation of the corresponding LCE. In the same section we establish a numerical criterion, that is the dFNVI, in order to quantify the results obtained by the FNVI method. In section 3 we demonstrate the ability of the new method to reveal the detailed structure of the dynamics in the phase space in both 2D and 3D dynamical systems. In the following section, we apply the new method to follow the evolution of a sticky orbit in the two-dimensional dynamical system. In section 5 we conduct a detailed comparison between the new FNVI/dFNVI method and with some other relatively modern well-known methods of chaos detection. Finally, in section 6 we summarize our results and we present a discussion and also the conclusions of the present research.

\section{The definition of the FNVI method}

The basic idea behind the FNVI method is the introduction of a simple and easily computed quantity that clearly identifies the ordered or chaotic nature of an orbit. The FNVI is defined as
\begin{equation}
\rm FNVI(t) = \frac{1}{t} \left| \frac{\|\vec{x}(t)\| - \|\vec{x}(0)\|}{\|\vec{x}(0)\|} \right|,  \ \ \ \ t \leq t_{max},
\end{equation}
where $t$ is the time, $\| \cdot \|$ denotes the Euclidean norm of the vector $\vec{x}(t)$, while $\|\vec{x}(0)\|$ and $\|\vec{x}(t)\|$ are the norm vectors for a given orbit, at times $t = 0$ and $t > 0$ respectively. In practice, we stop the evolution of $\vec{x}(t)$ after some time $t_1 = t^{\ast}$, we record the computed $\rm FNVI(t_1)$ and then we repeat the calculation for another time interval $t_2 = t^{\ast}$, etc obtaining finally the FNVI as a summation over many $t_i$, $i = 1, 2, . . . , N$. Here we must point out, that $t^{\ast}$ is the predefined time step of the numerical integration which remains constant during the entire predefined total integration time $t_{max}$.

In order to apply the FNVI method, we shall investigate the nature of orbits in a dynamical system of perturbed harmonic oscillators given by the 3D potential
\begin{equation}
V(x,y,z) = \frac{\omega^2}{2} \left(x^2 + y^2 + z^2 \right) + \epsilon \left(x^2y^2 + y^2z^2 + x^2z^2 \right),
\end{equation}
where $\omega$ is the common frequency of the oscillations along the $x$, $y$ and $z$ axis, while $\epsilon$ is the strength of the perturbation. Potential (5) represents three coupled harmonic oscillators in the case of the 1:1:1 resonance. Potentials of this type are also known as perturbed elliptic oscillators [1, 3, 7]. The basic reason for the choice of potential (5) is that perturbed elliptic oscillators appear very often in galactic dynamics and also in atomic-particle physics [7]. A second reason is that is displays exact periodic orbits, interesting sticky orbits together with large unified chaotic domains. Therefore, it gives us a great opportunity to test and prove the effectiveness and the reliability of the new FNVI method.

The Hamiltonian function to the potential (5) reads
\begin{eqnarray}
H_3\left(x, y, z, p_x, p_y, p_z \right) &=& \frac{1}{2} \left(p_x^2 + p_y^2 + p_z^2 \right) + V(x,y,z) \nonumber \\
 &=& h_3,
\end{eqnarray}
where $p_x$, $p_y$ and $p_z$ are the momenta per unit mass conjugate to $x$, $y$ and $z$ respectively, while $h_3$ is the numerical of the Hamiltonian.

The corresponding Hamiltonian for the 2D system can easily be obtained if we set $z = p_z = 0$ in equation (6). Then
\begin{eqnarray}
H_2\left(x, y, p_x, p_y \right) &=& \frac{1}{2} \left(p_x^2 + p_y^2 \right) + V(x,y) \nonumber \\
    &=& \frac{1}{2} \left(p_x^2 + p_y^2 + \omega^2x^2 + \omega^2y^2 \right) + \epsilon x^2y^2 \nonumber \\
    &=& h_2,
\end{eqnarray}
where $h_2$ is the numerical value of the 2D Hamiltonian.

The outcomes of the present research are mainly based on the numerical integration of the equations of motion
\begin{eqnarray}
\ddot{x} &=& - \frac{\partial V(x,y,z)}{\partial x} =
- \left[\omega^2 + 2\epsilon \left(y^2 + z^2 \right) \right] x, \nonumber \\
\ddot{y} &=& - \frac{\partial V(x,y,z)}{\partial y} =
- \left[\omega^2 + 2\epsilon \left(x^2 + z^2 \right) \right] y, \nonumber \\
\ddot{z} &=& - \frac{\partial V(x,y,z)}{\partial z} =
- \left[\omega^2 + 2\epsilon \left(x^2 + y^2 \right) \right] z,
\end{eqnarray}
where the dot indicates derivative with respect to the time. We integrated numerically the equations of motion with a double precision Bulirsch-St\"{o}er algorithm in FORTRAN 95. The accuracy of our results was checked by the constancy of the energy integrals (6) and (7), which were conserved up to the fifteenth significant decimal point.

We keep the involved parameters of the two systems fixed at the values $\omega = 1$ and $\epsilon = 1$, while the value of the energy $h_2$ or $h_3$ is treated as a parameter.

A simple qualitative way of studying the dynamics of a Hamiltonian system is by plotting the successive intersections of the orbits with a Poincar\'{e} surface of section [17]. This method has been extensively applied to 2D Hamiltonians, as in these systems the PSS is a two-dimensional plane. In 3D systems, however, the PSS is four dimensional and the behavior of the orbits cannot be easily visualized. One way to overcome this problem is to project the PSS to spaces with lower dimensions (see, [31, 32]). However, even these projections are often very complicated and difficult to be interpreted.
\begin{figure*}[!tH]
\centering
\resizebox{\hsize}{!}{\rotatebox{0}{\includegraphics*{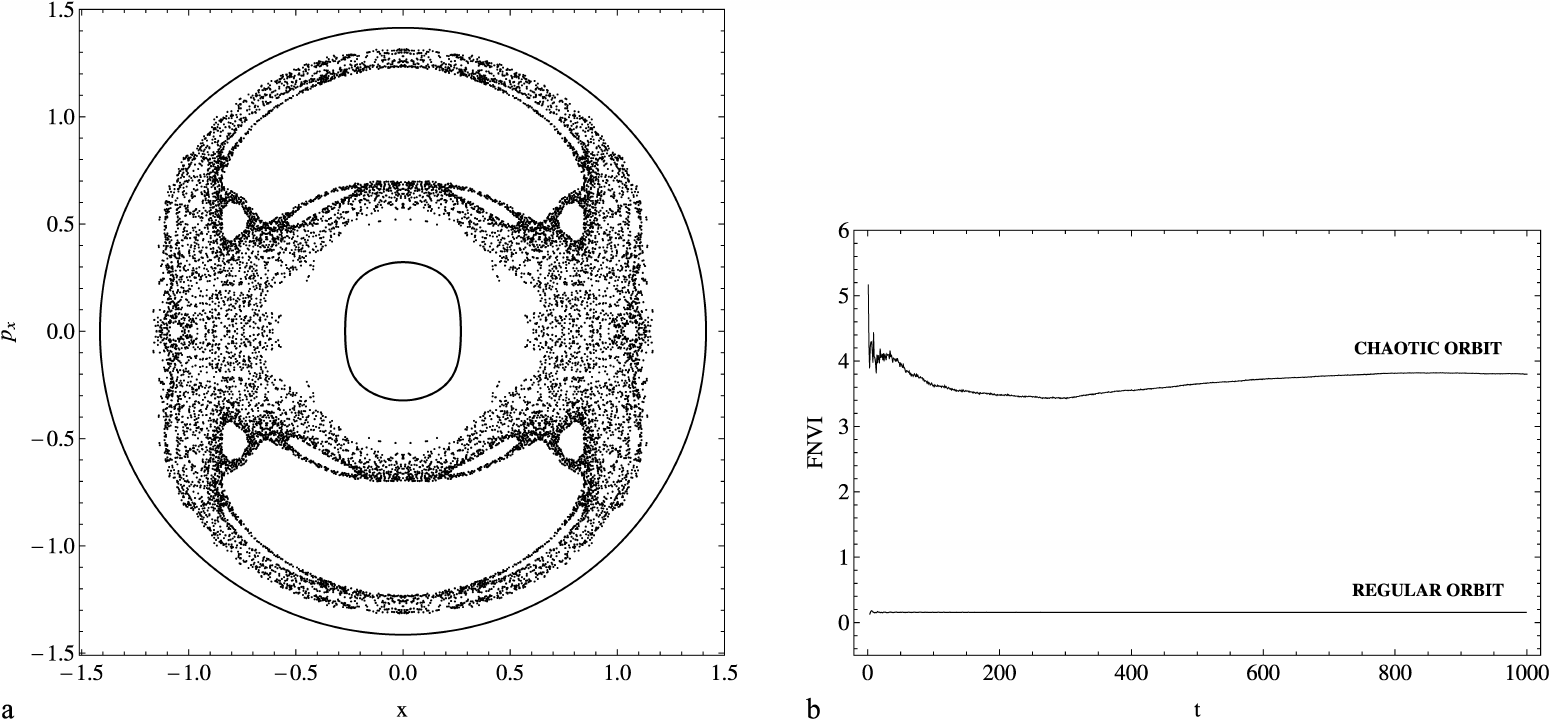}}}
\vskip 0.01cm
\caption{(a-b): (a-left): The PSS of an ordered and a chaotic orbit of the Hamiltonian system (7). The ordered orbit corresponds to a closed (solid) elliptic curve, while the chaotic one is represented by the dots scattered over the PSS. The outermost black solid line is the ZVC. (b-right): The time evolution of the FNVI for the two orbits of panel (a).}
\end{figure*}

In order to illustrate the behavior of the FNVI in 2D and 3D dynamical systems, we first consider some representative ordered and chaotic orbits. In Figure 1a we plot the intersection points of an ordered and a chaotic orbit of the dynamical system (7), with a $(x, p_x)$ PSS defined by $y = 0$ and $p_y >0$. The points of the ordered orbit lie on a torus and form a smooth closed curve on the PSS. On the other hand, the points of the chaotic orbit appear randomly scattered. The outermost black solid line shown in Fig. 1a is the Zero Velocity Curve (ZVC). The equation of the limiting curve — ZVC (that it the curve containing
all the invariant curves of the 2D system) is defined by the equation
\begin{equation}
f_2\left(x, p_x \right) = \frac{1}{2} p_x^2 + V(x) = h_2.
\end{equation}
The time evolution of the FNVI for these two orbits and for a time period of $10^3$ time units is plotted in Figure 1b. In the case of the ordered orbit the FNVI remains almost constant, while in the case of the chaotic orbit it displays large fluctuations. The initial conditions for the regular orbit are: $x_0 = 0.27$, $y_0 = 0$ and $p_{x0} = 0$, while the initial value of $p_{y0}$ is found from the energy integral (7). The chaotic orbit has initial conditions: $x_0 = 0.82$, $y_0 = 0$ and $p_{x0} = 0$. For both orbits the value of the energy is $h_2 = 1$, while the the PSS shown in Fig. 1a was integrated for a time period of 5 $\times$ $ 10^3$ time units. In order to have a better and more closer view of the evolution of the FNVI for these two orbits, we present separately in Figure 2a-b the two diagrams. We observe, that in Fig. 2a which corresponds to the regular orbit, the FNVI after a small transient period of fluctuation for about 200 time units, it stabilizes at a value and remains almost constant. On the contrary, in the case of the chaotic orbit shown in Fig. 2b the FNVI displays a completely different character with large and abrupt fluctuations throughout the entire period of the numerical integration.
\begin{figure*}[!tH]
\centering
\resizebox{\hsize}{!}{\rotatebox{0}{\includegraphics*{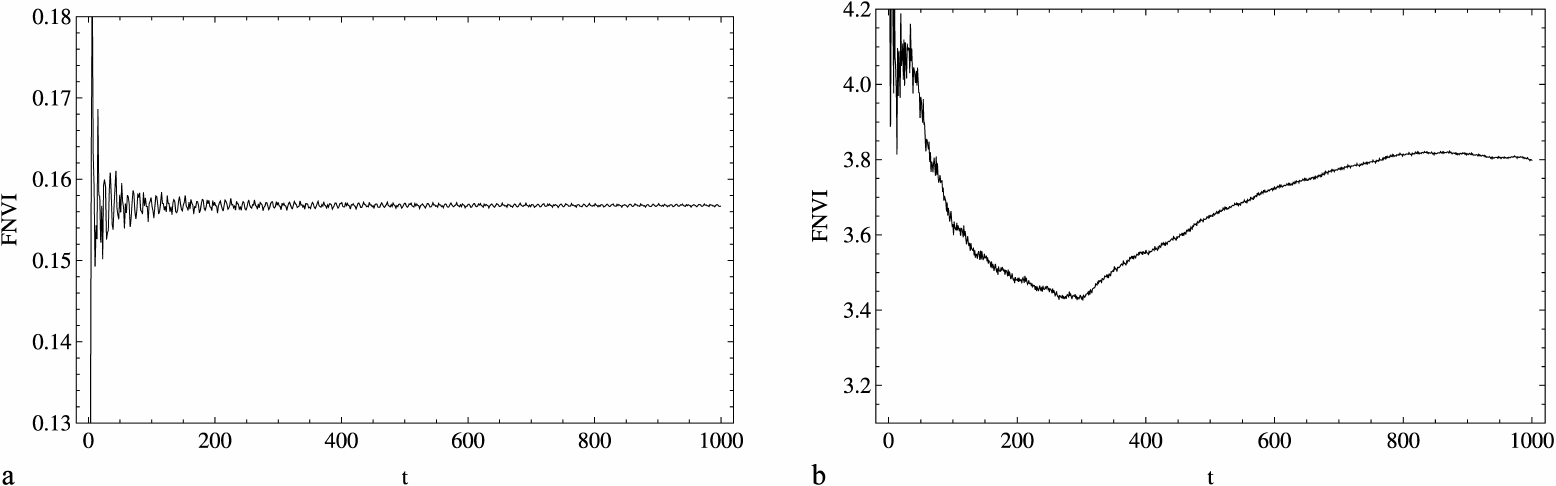}}}
\vskip 0.01cm
\caption{(a-b): The time evolution of the FNVI for a time interval of $10^3$ time units for a (a-left): regular 2D orbit and (b-right): chaotic 2D orbit. The initial conditions of the orbits and more details are given in the text.}
\end{figure*}
\begin{figure*}
\centering
\resizebox{\hsize}{!}{\rotatebox{0}{\includegraphics*{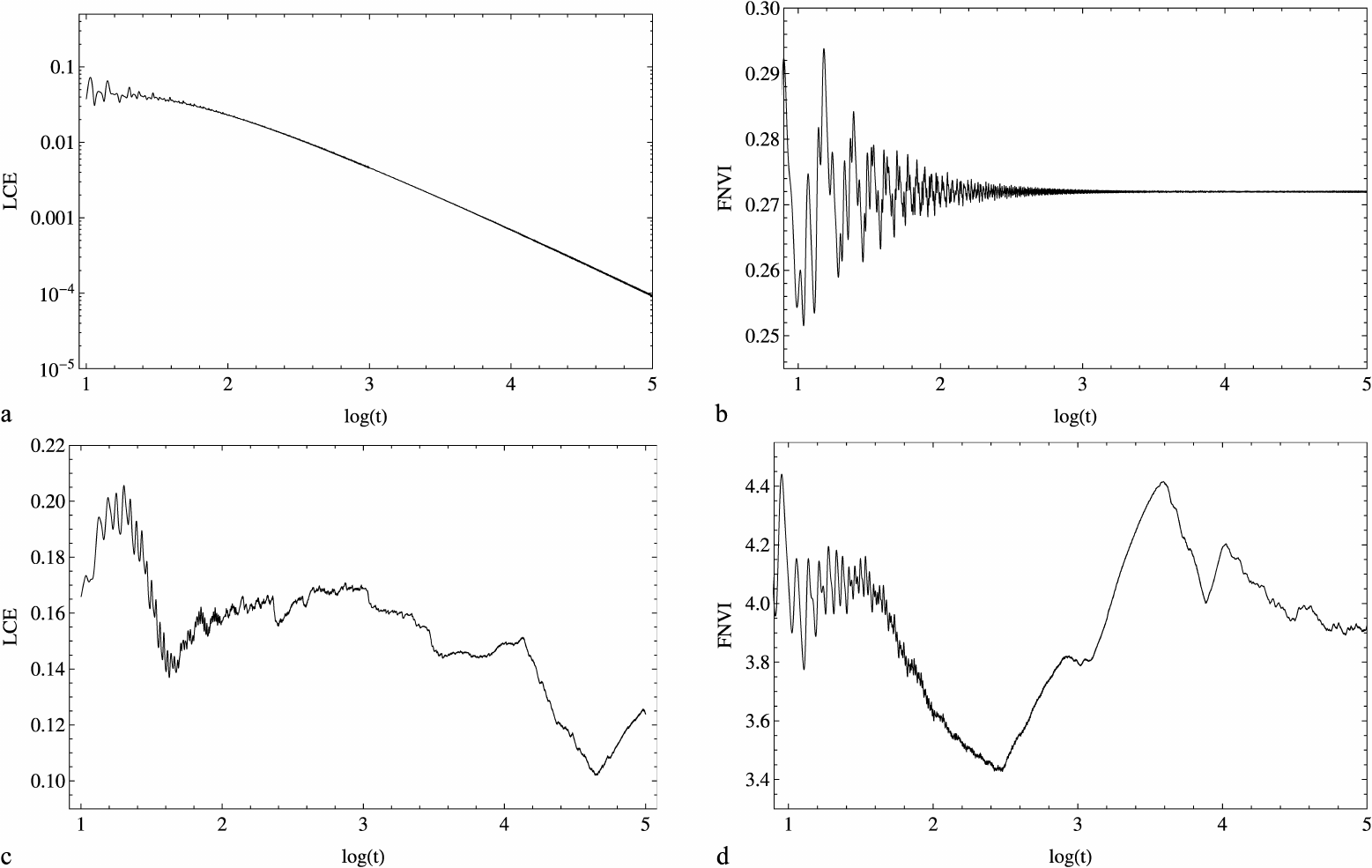}}}
\vskip 0.01cm
\caption{(a-d): The time evolution for a time interval of $10^5$ time units of (a-upper left): the LCE of the regular 2D orbit, (b-upper right): the FNVI of the same regular orbit, (c-lower left): the LCE of the chaotic 2D orbit and (d-lower right): the FNVI for the same chaotic orbit. Note that the $t$ axis is in log scale.}
\end{figure*}

One may reasonably assume that the shape of the FNVI depends on the particular value of the total time of the numerical integration. Thus, if we integrate the FNVI of the chaotic orbit shown in Fig. 2b for a larger time period it might displays a similar pattern as the one of the ordered orbit shown in Fig. 2a. In order to clarify this issue, we computed the FNVI for the same two 2D orbits discussed in Fig. 2a-b but for a time period of $10^5$ time units. The results are presented in Figs. 3b and 3d respectively and compared with the corresponding LCE in order to test their validity. In Figure 3a we see the evolution of the LCE of the ordered orbit. As expected the value of the LCE tends to zero as the time evolves. The evolution of the FNVI is presented in Figure 3b. Again, the FNVI after a small transient period of fluctuation for about 200 time units, it stabilizes at a value and remains almost constant. This indicates the regular character of the orbit. On the other hand, the computation of the maximal LCE, using Eqs. (1) and (2), despite its usefulness in many cases, does not have the same convergence properties over the same time interval. This becomes evident in Figure 3c where we plot the evolution of the LCE for the chaotic orbit. The computation of the LCE up to $t = 10^4$ or even up to $t = 10^5$ time units, still shows no clear evidence of convergence. Although Fig. 3c suggests that the orbit might probably be chaotic since LCE remains different from zero for large time intervals, it does not allow us to conclude its chaotic nature with certainty, and so further computation of the LCE is needed. Thus, it becomes evident that an advantage of the FNVI, with respect to the computation of the LCE, is that the current value of the FNVI is sufficient to determine the chaotic nature of an orbit, in contrast to the maximal LCE, where the whole evolution of the deviation vector affects the computed value of LCE. Figure 3d depicts the evolution of the FNVI for the chaotic orbit. Therefore, it becomes clear that the shape of the FNVI does not depend on the time of the numerical integration. We observe in Fig. 3d that the FNVI displays once more large and abrupt fluctuations throughout the time evolution. Note that in Figs. 3a-d the horizontal axis corresponding to the time is on log scale, in order to visualize more clearly the evolution of the indicators throughout the entire time interval. As the criterion by which we judge using the FNVI method, whether an orbit is regular or chaotic is qualitative rather than quantitative, we could be sure for the chaotic nature of the orbit after only $10^3$ time units of numerical integration.
\begin{figure*}
\centering
\resizebox{\hsize}{!}{\rotatebox{0}{\includegraphics*{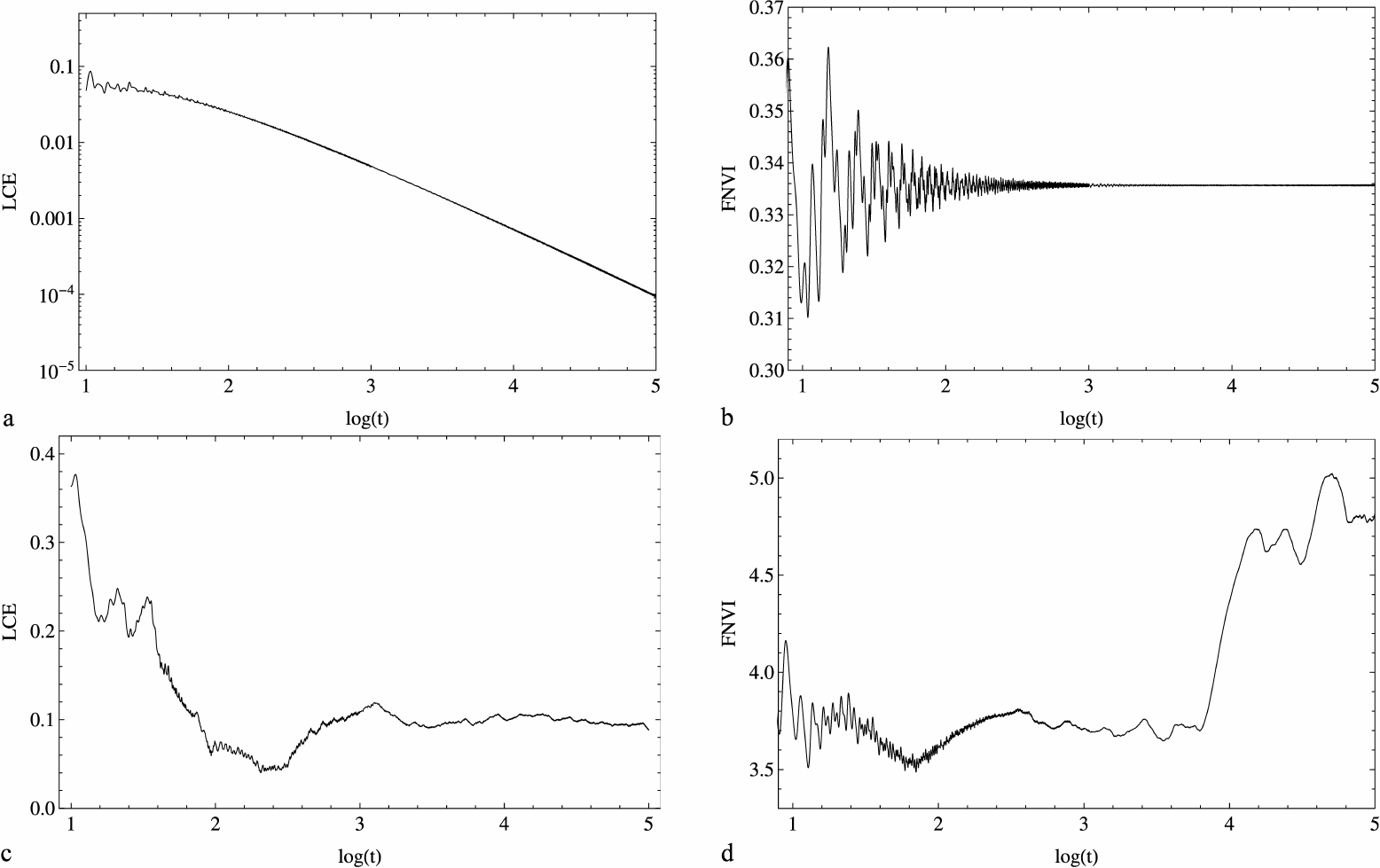}}}
\vskip 0.01cm
\caption{(a-d): The time evolution for a time interval of $10^5$ time units of (a-upper left): the LCE of a regular 3D orbit, (b-upper right): the FNVI of the same regular orbit, (c-lower left): the LCE of a chaotic 3D orbit and (d-lower right): the FNVI for the same chaotic orbit. Note that the $t$ axis is in log scale. The initial conditions of the orbits and more details are given in the text.}
\end{figure*}

Let us now present some results about the 3D Hamiltonian system (6). In Figure 4a we observe the evolution of the LCE of an ordered 3D orbit with initial conditions: $x_0 = 0.09, y_0 = 0, z_0 = 0.1, p_{x0} = p_{z0} = 0$, while the initial value of $p_{y0}$ is found from the energy integral (6). As expected, the value of the LCE tends to zero as the time evolves. The evolution of the FNVI is presented in Figure 4b. The FNVI after a small transient period of fluctuation for about 250 time units, it stabilizes at a value and remains almost constant, indicating the regular character of the orbit. In Figure 4c we see the evolution of the LCE of a chaotic 3D orbit. In this case the initial conditions of the chaotic orbit are: $x_0 = 0.86, y_0 = 0, z_0 = 0.1, p_{x0} = p_{z0} = 0$. The LCE remains different from zero, for a time period of $10^5$ time units, which implies the chaotic nature of the orbit. The corresponding evolution of the FNVI is presented in Figure 4d. It is clear that the pattern presented in Fig. 4d is completely different from that shown in Fig. 4b, where the 3D orbit is regular. For the 3D chaotic orbit, the FNVI displays large fluctuations and a highly irregular shape. Note that the $t$ axis in Figs. 4a-d is in log scale. For both 3D orbits the value of the energy is $h_3 = 1$.

Here, we have to point out that the criterion by which we determine so far the regular or chaotic nature of an orbit (2D or 3D), using the FNVI method is purely qualitative. In the case of a regular orbit the FNVI after a small transient period of fluctuation it stabilizes at a value and remains almost constant. On the other hand, when the orbit is chaotic the evolution of the corresponding FNVI follows a completely different pattern. It does not shows any evidence of convergence and displays large and abrupt fluctuations. This criterion has been obtained by testing a large number of regular and chaotic orbits in both dynamical systems (2D and 3D). Of course, since our criterion is qualitative we have to inspect the shape of FNVI each time in order to characterize an orbit. Obviously, this is not very practical when someone wants to check a large volume of orbits, so as to form an idea about the global structure of the dynamical system. Thus, we need to establish a new numerical criterion, in order to quantify the results obtained by the FNVI method. This criterion can be derived by looking the shape of the FNVI shown in Figs. 3b, 3d, 4b and 4d. One may observe, that when the orbit is regular the FNVI remains almost constant, while in the case of a chaotic orbit it displays high fluctuations. We are going to exploit this significant difference in order to obtain our quantitative criterion. Thus, we calculate the maximum and the minimum value of FNVI when $t \in [200, 1000]$. We take this particular time interval because in the case where the orbit is regular, for $t \lesssim 200$ time units there is a transient period of fluctuation, which may create a malfunction to our criterion. Using the above procedure we can define the
\begin{equation}
\rm dFNVI = FNVI_{max} - FNVI_{min},
\end{equation}
where $\rm FNVI_{max}$ and $\rm FNVI_{min}$ are the maximum and the minimum value of FNVI respectively when $t \in [200, 1000]$. The value of dFNVI can provide us the quantitative criterion that we seek. We note, that it is not easy to define a threshold value, so that the dFNVI being larger than this value reliably signifies chaoticity. Nevertheless, extensive numerical experiments in both dynamical systems indicate that in general, a good guess for this value could be 0.05. Thus, when dFNVI $\geqslant$ 0.05 the orbit is chaotic, while when dFNVI $<$ 0.05 the orbit ir ordered. This threshold value applies in both 2D and 3D orbits. For the 2D regular orbit presented in Fig. 2a dFNVI = 0.009, while for the 2D chaotic orbit of Fig. 2b dFNVI = 0.45. Similarly, for the 3D regular orbit presented in Fig. 4b dFNVI = 0.006, while for the 3D chaotic orbit of Fig. 4d dFNVI = 0.52. Therefore, we may conclude that the quantification of the FNVI method has been archived successfully. In the next section, we shall provide more extensive and detailed results using the dFNVI.

\section{Distinguishing between regions of order and chaos}

The dFNVI offers indeed a very easy and efficient method for distinguishing the chaotic versus ordered nature of orbits in a variety of problems. In the present section, we use it for identifying regions of phase space where large scale ordered and chaotic motion are both present. In particular, we shall apply the dFNVI method in order to reveal the structure in both Hamiltonian systems of two (2D) and three (3D) degrees of freedom.
\begin{figure*}[!tH]
\centering
\resizebox{\hsize}{!}{\rotatebox{0}{\includegraphics*{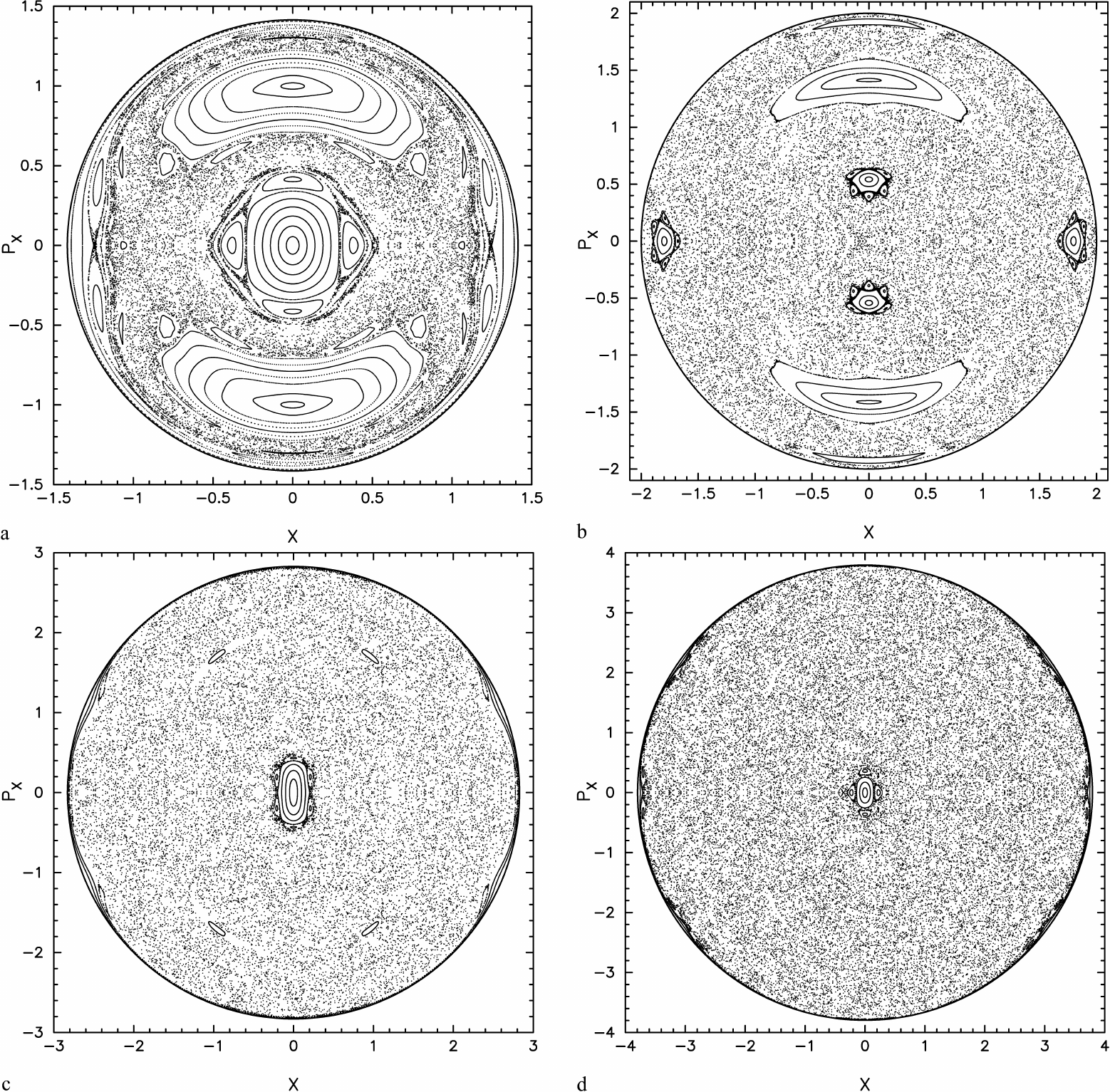}}}
\vskip 0.01cm
\caption{(a-d): The $(x, p_x)$, $y = 0, p_y > 0$ phase plane for the 2D Hamiltonian system (7), when (a-upper left): $h_2 = 1$, (b-upper right): $h_2 = 2$, (c-lower left): $h_2 = 4$ and (d-lower right): $h_2 = 7.2$. More details are provided in the text.}
\end{figure*}

\subsection{Order and chaos in the 2D dynamical system}

In Figure 5a-d we present detailed plots of the $(x,p_x)$, $y = 0$, $p_y > 0$ PSS of the 2D dynamical system (7) for different values of the energy $h_2$. Fig. 5a shows the PSS plot, when $h_2 = 1$. One can see, that regions of ordered motion around stable periodic orbits are seen to coexist with chaotic regions filled by scattered points. In particular, a large part of the phase plane is covered by chaotic orbits, while the regular regions are occupied by invariant curves mainly around the points $(x_0, p_{x0}) = (0, 0)$ and $(x_0, p_{x0}) = (0, \pm \sqrt{h_2})$. The above points give the position of two exact periodic orbits. The first is the $y$ axis, while the second describes the $x = \pm$ $y$ straight-line orbits on the $(x, p_x)$ phase plane. Furthermore, one observes a large number of smaller islands of invariant curves produced by secondary resonances and some sticky regions as well. Fig. 5b is similar to Fig. 5a but when $h_2 = 2$. Here, we see that the majority of the phase plane is covered by a unified chaotic sea. The regular regions are confined only around the straight line periodic orbits, while the $y$ axis has now become unstable. Some smaller islands of invariant curves are also present. The most interesting feature observed in this case, is the sticky regions around the chain of islands of invariant curves on the $x$ as well as on the $p_x$ axis. If we increase the value of energy to $h_2 = 4$, we obtain the phase plane shown in Fig. 5c. In this case, the $y$ axis has returned to stability, while the straight line periodic orbits are unstable. Almost all the phase plane is chaotic, except of a small region around the center and some small islands of invariant curves which are products of secondary resonant orbits. A sticky region around the chain of the small islands of invariant curves near the center is also observed. Fig. 5d is similar to Fig. 5a, but when $h_2 = 7.2$. Here, almost the entire phase plane is covered by chaotic orbits. There is a very small regular region confined near the center. A careful observation also reveals some very tiny regular islands of invariant curves embedded in the vast chaotic sea.
\begin{figure*}
\centering
\resizebox{\hsize}{!}{\rotatebox{0}{\includegraphics*{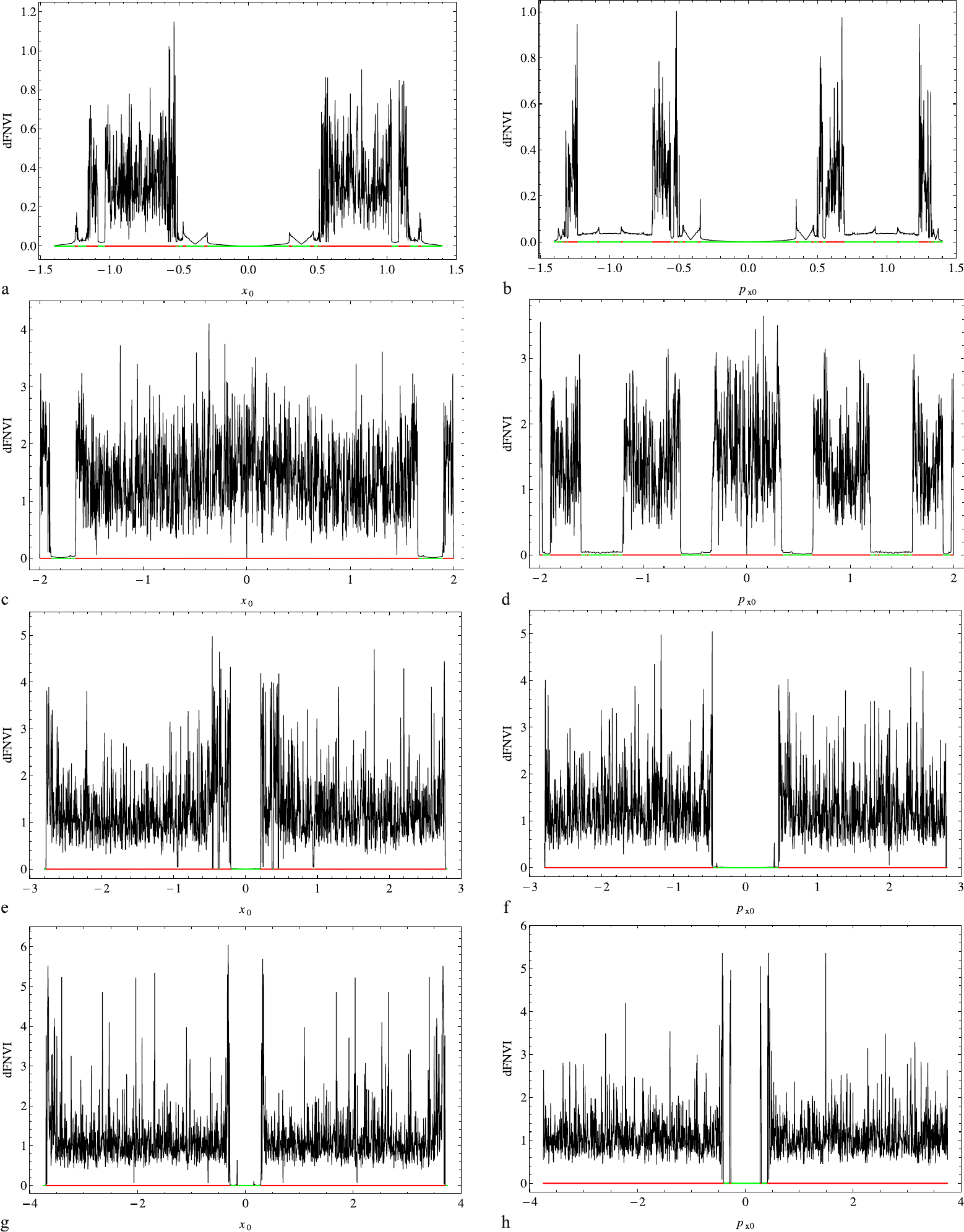}}}
\vskip 0.01cm
\caption{(a-h): The values of the dFNVI for orbits of the 2D system (7) with initial conditions on the (left pattern): $p_x = 0$ line as a function of the $x_0$ coordinate of the initial conditions $(x_0, 0)$, (right pattern): $x = 0$ line as a function of the $p_{x0}$ coordinate of the initial conditions $(0, p_{x0})$, for the PSS plots of Fig. 5a-d. Green and red dots indicate initial conditions corresponding to regular and chaotic orbits respectively. See text for more details.}
\end{figure*}

In order to demonstrate and prove the effectiveness of the dFNVI method, we first consider orbits whose initial conditions lie on the lines $x = 0$ and $p_x = 0$. In particular, we take 5 $\times$ $10^3$ equally spaced initial conditions on these lines and we compute the value of the dFNVI for each one. The results are presented in Figure 6a-h where we plot the dFNVI as a function of the $x_0$ and $p_{x0}$ coordinates of the initial conditions of these orbits for $t = 10^3$ time units. In all panels the data points are line connected, so that the changes of the dFNVI values are clearly visible. With green dots we represent the initial conditions $(x_0, 0)$ or $(0, p_{x0})$ corresponding to regular orbits, while the initial conditions producing chaotic orbits are marked with red dots. Note that there are intervals where the dFNVI has low values (e.g. lower than 0.05), which correspond to ordered motion inside the islands of stability crossed by the $x = 0$ and $p_x = 0$ lines shown in Fig. 5a-d. There also exist regions where the dFNVI has large values (e.g. larger than 0.1) denoting that in these regions the motion is chaotic. These intervals correspond to the regions of scattered points crossed by the $x = 0$ and $p_x = 0$ lines in Fig. 5a-d. Although most of the initial conditions give large $(> 0.1)$ or very small $(< 10^{-2})$ values for the dFNVI, there also exist initial conditions that have intermediate values of the dFNVI (0.05 $<$ dFNVI $<$ 0.1). These initial conditions correspond to sticky chaotic orbits, remaining for long time intervals at the borders of islands, whose chaotic nature will be revealed later on. In Figs. 6a and 6b we observe the values of the dFNVI computed for $t = 10^3$ time units for orbits with initial conditions on the $p_x = 0$ and $x = 0$ lines on the PSS shown in Fig. 5a, as a function of the $x_0$ coordinate and the $p_{x0}$ momenta respectively. Similarly, Figs. 6c and 6d correspond to the PSS of Fig. 5b, Figs. 6e and 6f correspond to the PSS of Fig. 5c and Figs. 6g and 6h correspond to the PSS shown in Fig. 5d.

In Fig. 6a, around $x_0 \approx 1.05$ there exists a group of points inside a large chaotic region having dFNVI $<$ 0.05. These points correspond to orbits with initial conditions inside a small stability island, which is not very visible in the detailed PSS plot of Fig. 5a. Also the points with $p_{x0} = \pm$ 0.91 in Fig. 6b, have been characterized by the dFNVI as initial conditions corresponding to chaotic orbits, while all their neighboring points have dFNVI $<$ 0.05. These points actually correspond to a weak chaotic separatrix inside the vast domain of stability, which can be revealed only after a very high magnification of this region of the PSS. So, we see that the systematic application of the dFNVI method can reveal very fine details of the structure in a dynamical system.

It will be of particular interest, to calculate the mean value of dFNVI for the regular and chaotic orbits discussed in Fig. 6a-h. In other words, we are going to calculate the mean value of dFNVI for the regular orbits, that is $\rm \langle dFNVI \rangle_R$ and the mean value of dFNVI for the chaotic orbits $\rm \langle dFNVI \rangle_C$, with initial conditions along the $x$ and the $p_x$ axis for different values of the energy $h_2$. In Table 1 we present our results. It is evident, that as the value of the energy increases, we have an increase at the mean value of the dFNVI for both regular and chaotic orbits. In the case of chaotic orbits, the increase of $\rm \langle dFNVI \rangle_C$ indicates that the chaoticity of the dynamical system increases as the value of the energy is amplified. We shall come back to this point later in this section.
\begin{table}[ht]
\caption{The mean value of dFNVI for the regular and chaotic orbits discussed in Fig. 6a-h.}
\centering
\begin{tabular}{|c||c|c|c|c|}
\hline \hline
 & \multicolumn{2}{c}{$x$ - axis} & \multicolumn{2}{c}{$p_x$ - axis} \bigstrut \\ \hline
$h_2$ & $\rm \langle dFNVI \rangle_R$ & $\rm \langle dFNVI \rangle_C$ & $\rm \langle dFNVI \rangle_R$ & $\rm \langle dFNVI \rangle_C$ \\
\hline \hline
1.0 & 0.00725 & 0.32251 & 0.00551 & 0.28687  \\
2.0 & 0.01136 & 1.14872 & 0.01922 & 1.15079  \\
4.0 & 0.01732 & 1.27763 & 0.02317 & 1.21792  \\
7.2 & 0.02181 & 1.43602 & 0.02761 & 1.42706  \\
\hline \hline
\end{tabular}
\end{table}
\begin{figure*}
\centering
\resizebox{\hsize}{!}{\rotatebox{0}{\includegraphics*{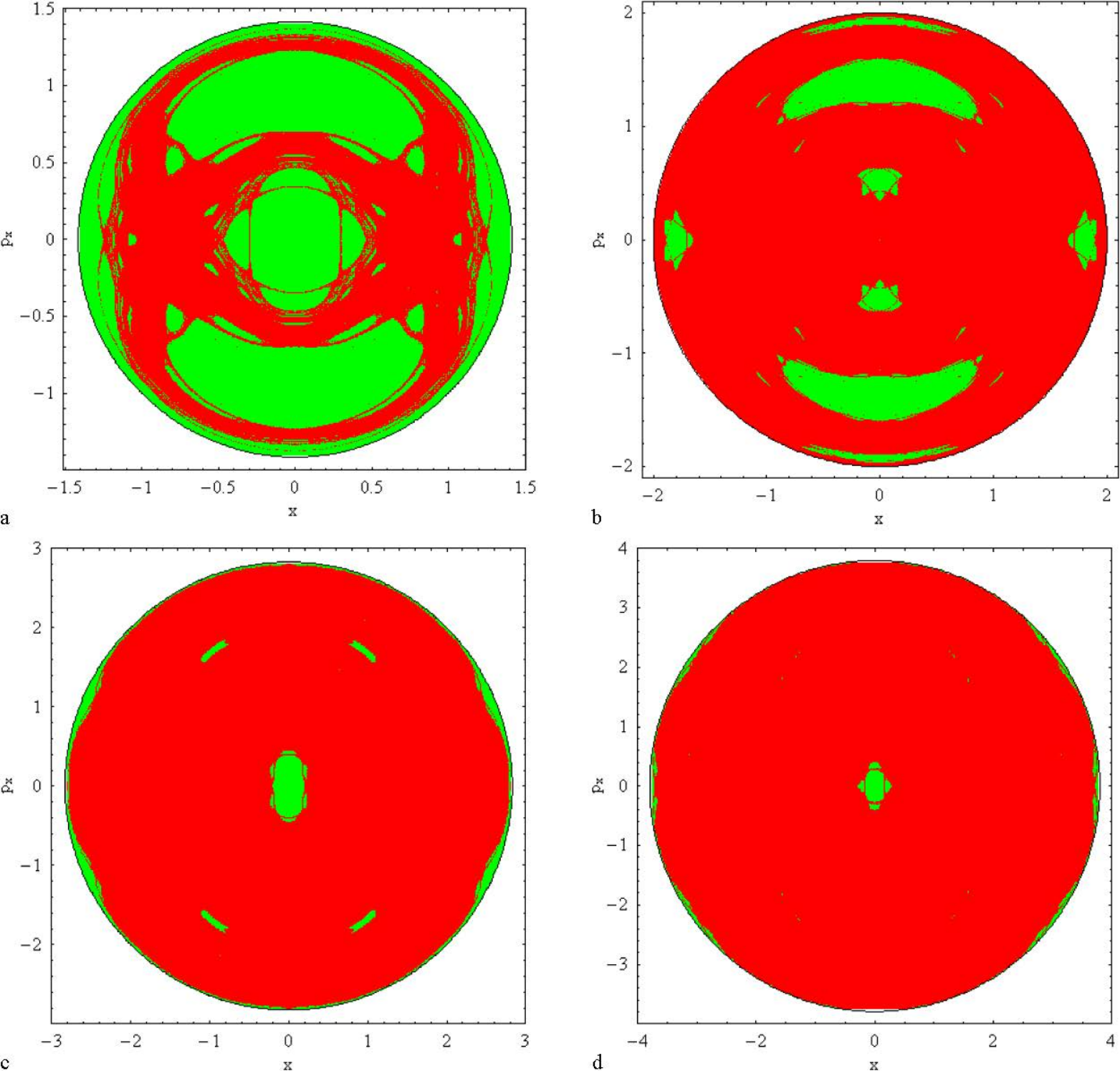}}}
\vskip 0.01cm
\caption{(a-d): Regions of different values of the dFNVI on the PSS plots of the 2D dynamical system (7) when (a-upper left): $h_2 = 1$, (b-upper right): $h_2 = 2$, (c-lower left): $h_2 = 4$ and (d-lower right): $h_2 = 7.2$. In all panels, the initial conditions $(x_0, p_{x0})$ are colored green if their dFNVI $<$ 0.05 and red if dFNVI $\geq$ 0.05.}
\end{figure*}

By carrying out the previously presented analysis for initial conditions $(x_0, p_{x0})$ not only along a line but on the whole phase plane of a PSS and giving to each point a color according to the value of the dFNVI, we can have a clear picture of the regions where chaotic or ordered motion occurs. The outcomes of this procedure for the 2D dynamical system (7), using a dense grid of initial conditions $\left(x_0, p_{x0}\right)$ on the PSS, are presented in Figure 7a-d. The value of the energy $h_2$ in Figs. 7a, 7b, 7c and 7d is the same as in Figs. 5a, 5b, 5c and 5d respectively. Thus, in Figs. 7a-d we clearly distinguish between green regions, where the motion is ordered and red regions, where it is chaotic. The outermost black solid line shown in Figs. 7a-d is the limiting curve (ZVC) defined by Eq. (9). It is worth mentioning, that in Fig. 7a we can observe small islands of stability inside the large chaotic sea, which are not very visible in the detailed PSS of Fig. 5a, such as that for $x_0 \approx \pm$ 1.05, $p_{x0} \approx $ 0. Although all the orbits with initial conditions $(x_0, p_{x0})$ in Figs. 7a-d were computed for only $t$ = $10^3$ time units (such as Figs. 6a-h), this time was sufficient enough for the clear revelation of small ordered regions inside the chaotic domains. We must point out, that the outcomes derived using the dFNVI method regarding the structure of the phase plane coincide with those obtained using the PSS technique. In other words, we see that the structure of the phase planes shown in Figs. 7a-d are practically identical with those presented in Figs. 5a-d. For a grid of about $10^5$ equally spaced initial conditions $\left(x_0, p_{x0}\right)$, we need about 7 h of CPU time on a Pentium Dual Core Processor at 2.2GHz PC, in order to construct each grid-plot shown in Figs. 7a-d.
\begin{figure}[!tH]
\centering
\resizebox{\hsize}{!}{\rotatebox{0}{\includegraphics*{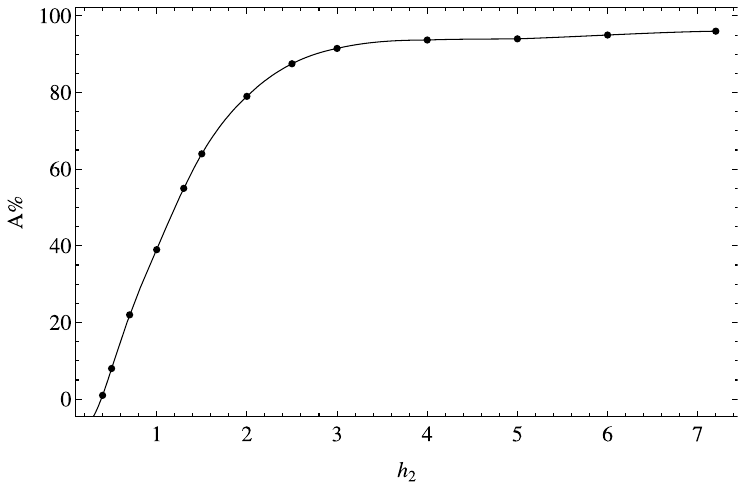}}}
\caption{A plot of the percentage of the phase plane $A\%$ covered by chaotic orbits versus the value of the energy $h_2$.}
\end{figure}

Grid-plots such as those of Fig. 7a-d, apart from presenting the regions of order and chaos, can also be used to provide very accurate estimations regarding the fraction of the phase-space volume occupied by chaotic or ordered orbits and also give us good initial guesses for the location of stable periodic orbits, in regions where the motion is ordered. One can conclude form Figs. 5a-d or from Figs. 7a-d, that the fraction of the phase plane covered by chaotic orbits increases as the value of the energy $h_2$ increases. Figure 8 shows a plot of the percentage of the phase plane $A\%$ covered by chaotic orbits versus $h_2$. We observe that the percentage $A\%$ increases rapidly, as the value of $h_2$ increases. Dots indicate the values of the percentage $A\%$ obtained numerically from the grid-plots, while the solid line is a fourth degree polynomial fitting curve. We must point out that the percentage $A\%$ is calculated as follows: when constructing the grid-plots, we count the initial conditions $\left(x_0, p_{x0}\right)$ corresponding to regular orbits and also the initial conditions $\left(x_0, p_{x0}\right)$ corresponding to chaotic orbits. Then, we divide the number of chaotic orbits by the total number of orbits and thus, we obtain the percentage $A\%$ for each value of the energy $h_2$. Note, that in every case the total number of the calculated orbits is about $10^5$, while the exact number of chaotic orbits varies depending on the value of the energy $h_2$.
\begin{figure*}
\centering
\resizebox{\hsize}{!}{\rotatebox{0}{\includegraphics*{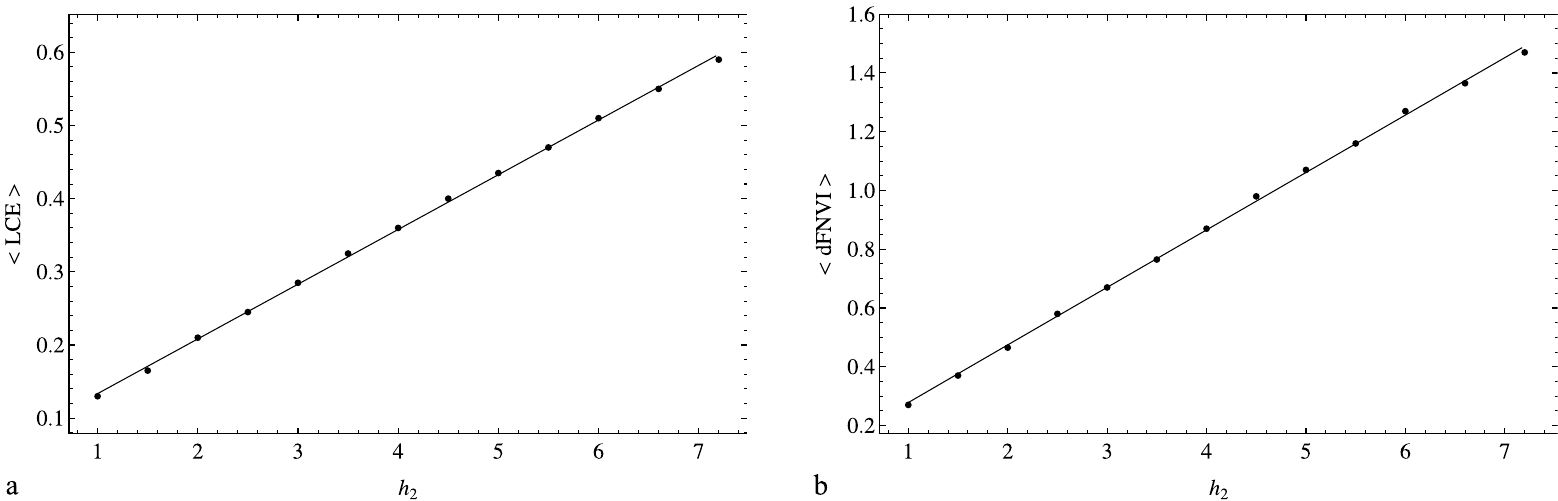}}}
\vskip 0.01cm
\caption{(a-d): (a-left): A plot of the average value of the maximal LCE versus $h_2$ and (b-right): A plot of the $\langle$ dFNVI $\rangle$ as a function of the value of the energy $h_2$.}
\end{figure*}

In order to have an estimation of the degree of chaos of the dynamical system from another point of view, we have plotted the average value of the maximal LCE versus $h_2$. The results are shown in Figure 9a. Note, that the $\langle$ LCE $\rangle$ increases linearly with $h_2$. Here, we must point out that it is well known that the value of the LCE is different in each chaotic component [22]. As we have in all cases regular regions and only one unified chaotic sea in each $(x, p_x)$ phase plane, we calculate the average value of LCE by taking $10^3$ orbits with different and random initial conditions $(x_0, p_{x0})$ in the chaotic domain for every value of the energy $h_2$. Note that all calculated LCEs are different on the fifth decimal point in the same chaotic region. We follow the same procedure using now the dFNVI. For the same $10^3$ initial conditions corresponding to chaotic orbits for each value of the energy $h_2$, we calculated the $\langle$ dFNVI $\rangle$. The results are presented in Figure 9b. Again the $\langle$ dFNVI $\rangle$ increases linearly with $h_2$. The increase of $\langle$ dFNVI $\rangle$ indicates that the chaoticity of the dynamical system increases as the value of the energy is amplified. As the behavior of $\langle$ dFNVI $\rangle$ coincides with the behavior of $\langle$ LCE $\rangle$, we prove once more the reliability of dFNVI method. However, the main advantage of the dFNVI method is that it requires only $10^3$ integration time units, while the LCE needs at least about 5 $\times$ $10^4$ integration time units so as to provide reliable and definitive evidence. Thus, the dFNVI is a much more faster indicator than the LCE.

\subsection{Order and chaos in the 3D dynamical system}

In the case of 3D Hamiltonian systems the PSS is now four dimensional and thus, not so useful as in the 2D systems. On the other hand, the dFNVI method has the ability once more to identify successfully regions of order and chaos in the phase space. To prove this, let us start with initial conditions on a 4D grid of the PSS. In this way, we find again regions of order and chaos, which may be visualized, if we restrict our study to a subspace of the whole 6D phase space. We consider orbits with initial conditions $(x_0, z_0, p_{x0})$, $y_0 = p_{z0} = 0$, while the initial value of $p_{y0}$ is always obtained from the energy integral (6). In particular, we define a value of $z_0$ which is kept constant and then we calculate the dFNVI of 3D orbits with initial conditions $(x_0, p_{x0})$, $y_0 = p_{z0} = 0$. Thus, we are able to construct again a 2D plot depicting the $(x, p_x)$ plane but with an additional value of $z_0$. All the initial conditions of the 3D orbits lie inside the limiting curve defined by
\begin{equation}
f_3\left(x, p_x; z_0 \right) = \frac{1}{2} p_x^2 + V(x;z_0) = h_3.
\end{equation}
\begin{figure*}
\centering
\resizebox{\hsize}{!}{\rotatebox{0}{\includegraphics*{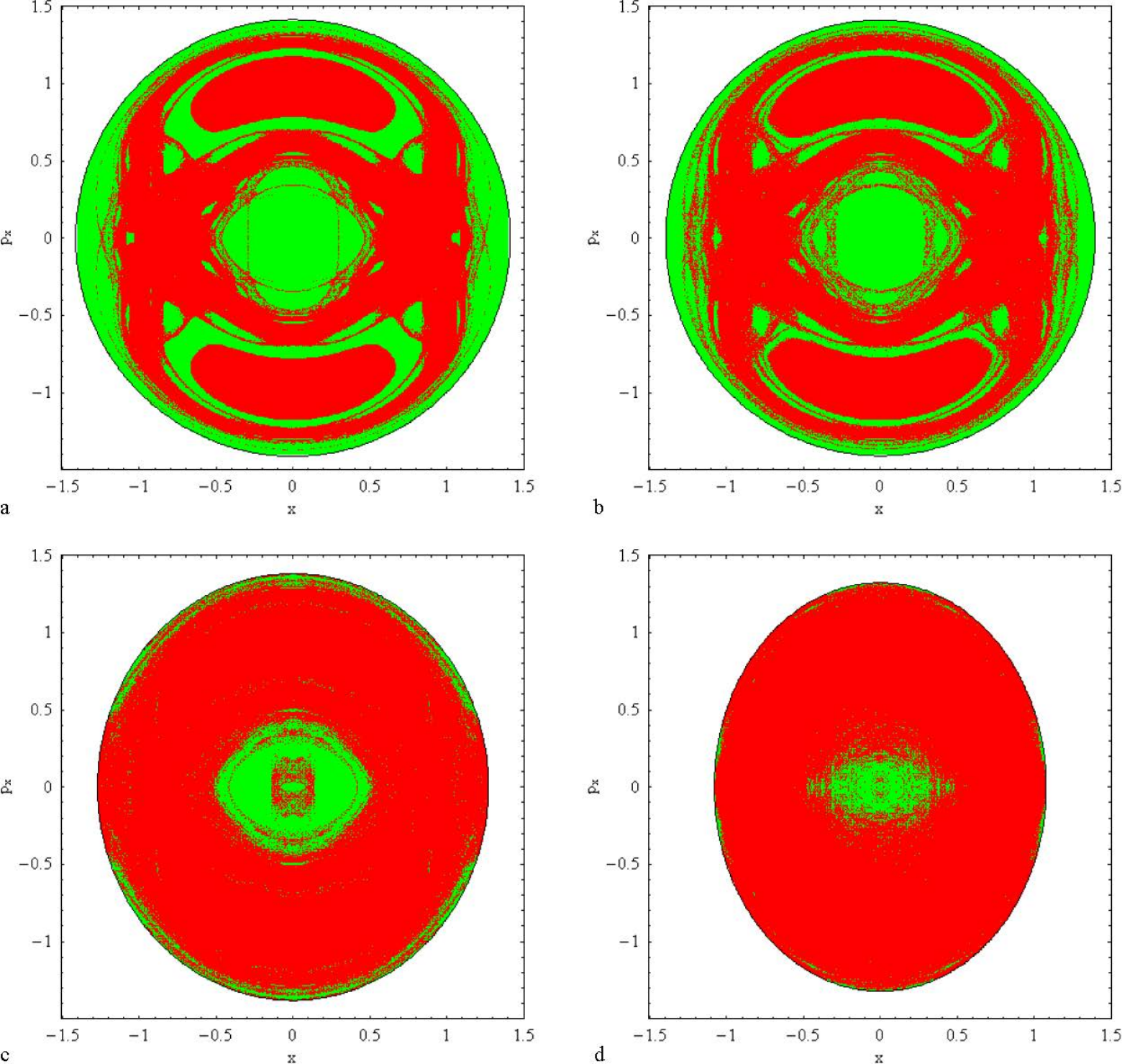}}}
\vskip 0.01cm
\caption{(a-d): Regions of different values of the dFNVI of the 3D dynamical system (6) for $h_3 = 1$, when (a-upper left): $z_0 = 0.01$, (b-upper right): $z_0 = 0.1$, (c-lower left): $z_0 = 0.3$ and (d-lower right): $z_0 = 0.5$. In all panels, the initial conditions $(x_0, z_0, p_{x0})$ are colored green if their dFNVI $<$ 0.05 and red if dFNVI $\geq$ 0.05.}
\end{figure*}

Our results are presented in Figure 10 a-d, where we give four grid-plots for the same value of the energy $h_3 = 1$, but for different initial value of $z_0$. Again, we can distinguish between regions of ordered (colored in green, where dFNVI $<$ 0.05) and chaotic motion (colored in red, where dFNVI $\geqslant$ 0.05). In Fig. 10a where $z_0 = 0.01$ we see that the structure of the $(x, p_x)$ plane is quite similar with the corresponding 2D grid-plot shown in Fig. 7a. We also see that well defined islands of stability inside the unified chaotic sea still exist. The main difference from the 2D grid-plot of Fig. 7a is that now in the 3D system, a large portion of orbits around the stable periodic points on the $p_x$ axis have altered their nature form ordered to chaotic. Fig. 10b is similar to Fig. 10a but when $z_0 = 0.1$. In this case, the structure of the $(x, p_x)$ plane is very similar to that shown in Fig. 10a. One may observe, that the percentage corresponds to chaotic orbits has been increased and moreover the islands of stability have begun to destabilize and lose their well defined structure. In Fig. 10c we increase the initial value of $z_0$ to 0.3. In this case, it is evident that the initial conditions correspond to ordered orbits are delocalized, since now we can see no signs of well defined islands of stability. Finally, in Fig. 10d we present a grid-plot when $z_0 = 0.5$. Here, the vast majority of the initial conditions correspond to chaotic orbits, while the initial conditions correspond to ordered orbits are completely delocalized and randomly scattered mainly in the central part of the $(x, p_x)$ plane. All the 3D orbits with initial conditions $(x_0, z_0, p_{x0})$ in Figs. 10a-d were computed for only $10^3$ time units. With a closer look to Figs. 10a-d one may conclude the following two main points: (i) the increase of the initial value of $z_0$ has as a result the increase of the percentage on the $(x, p_x)$ plane corresponding to chaotic orbits and (ii) as we amplify the value of $z_0$ we observe that the entire area of the $(x, p_x)$ plane defined by Eq. (11) is reduced and becomes smaller. This can easily be explained. The coordinates $(x, y, z)$ and the momenta $(p_x, p_y, p_z)$ of the 3D orbits are connected through the energy integral (6) and since we increase the value of $z_0$ while keeping constant the value of the energy $h_3$ this has as the result the permissible values of the initial conditions $(x_0, p_{x0})$ to be reduced. For a grid of about $10^5$ equally spaced initial conditions $\left(x_0, p_{x0}\right)$, we need about 7.5 h of CPU time on a Pentium Dual Core Processor at 2.2GHz PC, in order to construct each grid-plot shown in Figs. 10a-d.
\begin{figure}[!tH]
\centering
\resizebox{\hsize}{!}{\rotatebox{0}{\includegraphics*{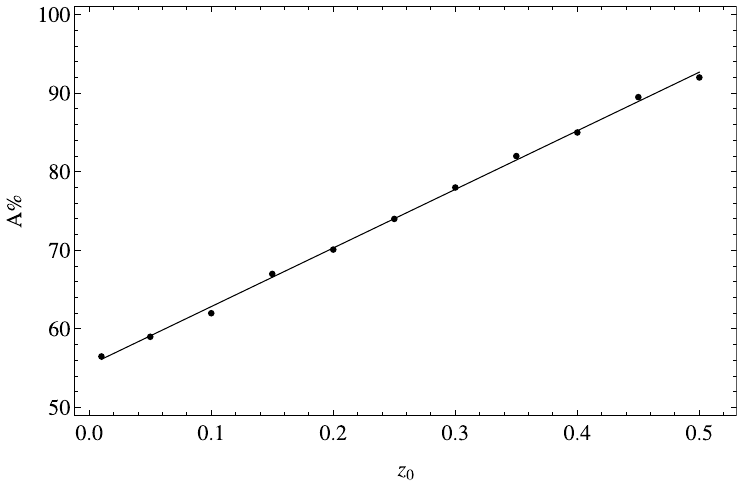}}}
\caption{A plot depicting the relationship between the percentage $A\%$ of the initial conditions $\left(x_0, p_{x0}\right)$ corresponding to chaotic 3D orbits and the initial value of $z_0$, when $h_3 = 1$.}
\end{figure}

The grid-plots shown in Figs. 10a-d could be considered as $\left(x, p_x\right)$ ``phase planes" but not with the strict definition. In fact, these plots show the structure of the $\left(x, p_x\right)$ subspace for a given value of the $z_0$. Therefore, we study the subspace $\left(x, p_x\right)$, $z = z_0$, $y = p_z = 0$, with $p_y > 0$ of the 6D phase space of the 3D dynamical system. Note that grid-plots such as those of Figs. 10a-d, apart from presenting the initial conditions correspond to ordered and chaotic orbits, can also be used to estimate the fraction of the phase-space volume occupied by chaotic or ordered 3D orbits. In Figure 11, we present a plot depicting the relationship between the percentage $A\%$ of initial conditions $\left(x_0, p_{x0}\right)$ corresponding to chaotic 3D orbits and the initial value of $z_0$. We observe that the chaotic percentage $z_0$ increases linearly with $z_0$. Our numerical calculation suggest, that when $z_0 \geq 0.67$ for $h_3 = 1$ eventually all the initial conditions $\left(x_0, z_0, p_{x0}\right)$ correspond to chaotic orbits, while regular motion, if any, is negligible. Note, that the plot shown in Fig. 11 corresponds to $h_3 = 1$. For different values of the energy the relationship between $A\%$ and $z_0$ is quite similar, but as we amplify the value of the energy $h_3$, the increase of the chaotic percentage is more rapid and does not follow a linear pattern any longer.
\begin{figure*}[!tH]
\centering
\resizebox{\hsize}{!}{\rotatebox{0}{\includegraphics*{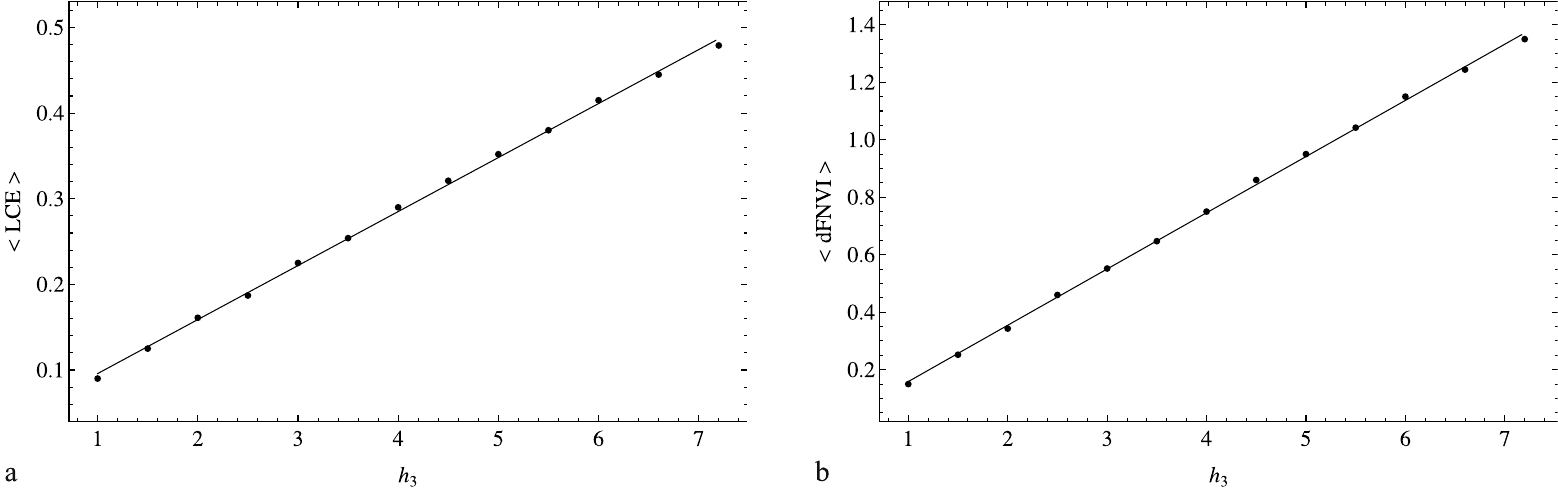}}}
\vskip 0.01cm
\caption{(a-d): (a-left): A plot of the average value of the maximal LCE versus $h_3$ and (b-right): A plot of the $\langle$ dFNVI $\rangle$ as a function of the value of the energy $h_3$, when $z_0 = 0.01$.}
\end{figure*}
\begin{figure*}[!tH]
\centering
\resizebox{\hsize}{!}{\rotatebox{0}{\includegraphics*{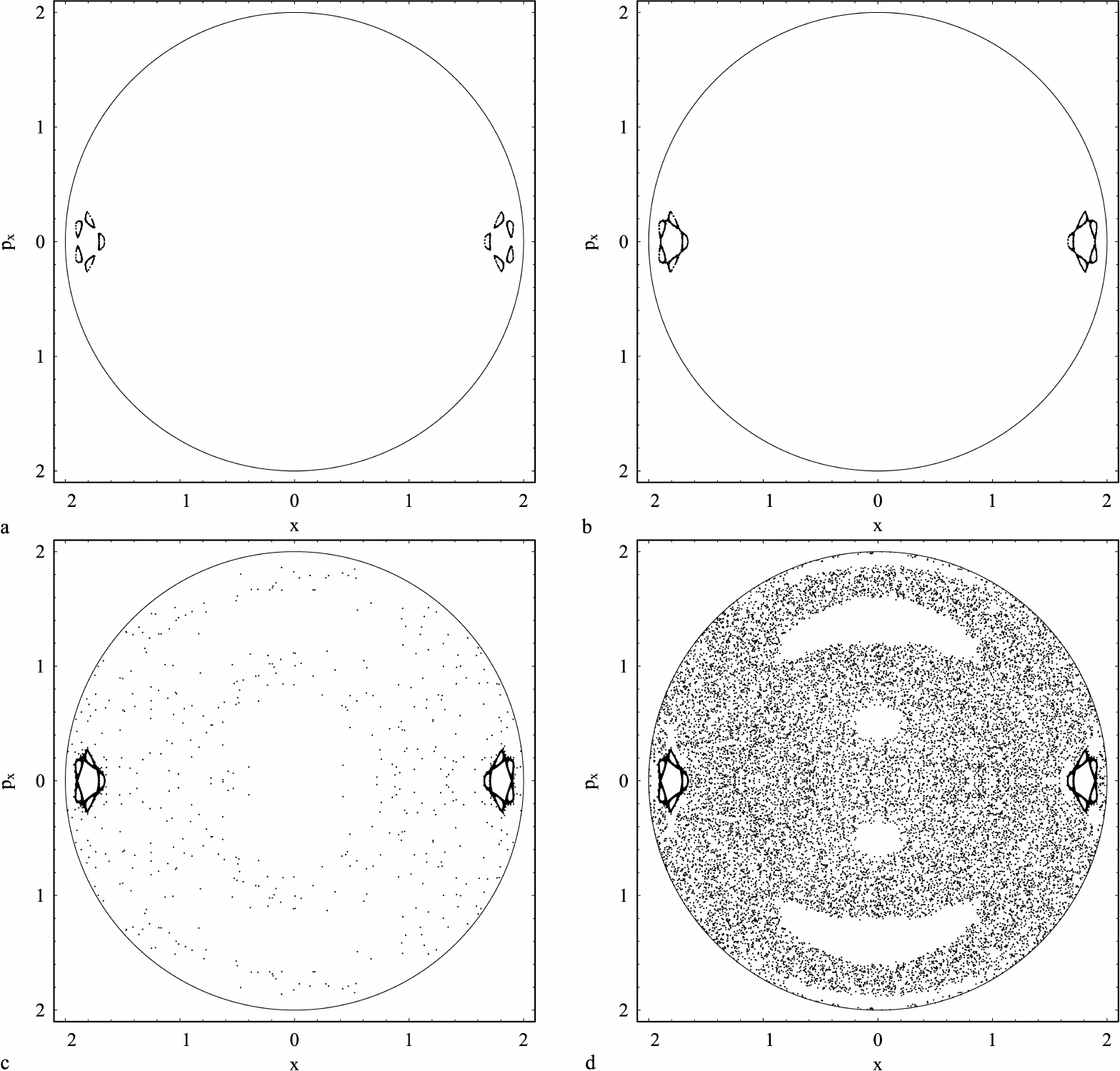}}}
\vskip 0.01cm
\caption{(a-d): Time evolution of the sticky orbit in the $(x, p_x)$ phase plane. The integration time is (a-upper left): $T = 1200$ time units, (b-upper right): $T = 1700$ time units, (c-lower left): $T = 12500$ time units and (d-lower right): 5 $\times$ $10^4$ time units.}
\end{figure*}
\begin{figure*}[!tH]
\centering
\resizebox{\hsize}{!}{\rotatebox{0}{\includegraphics*{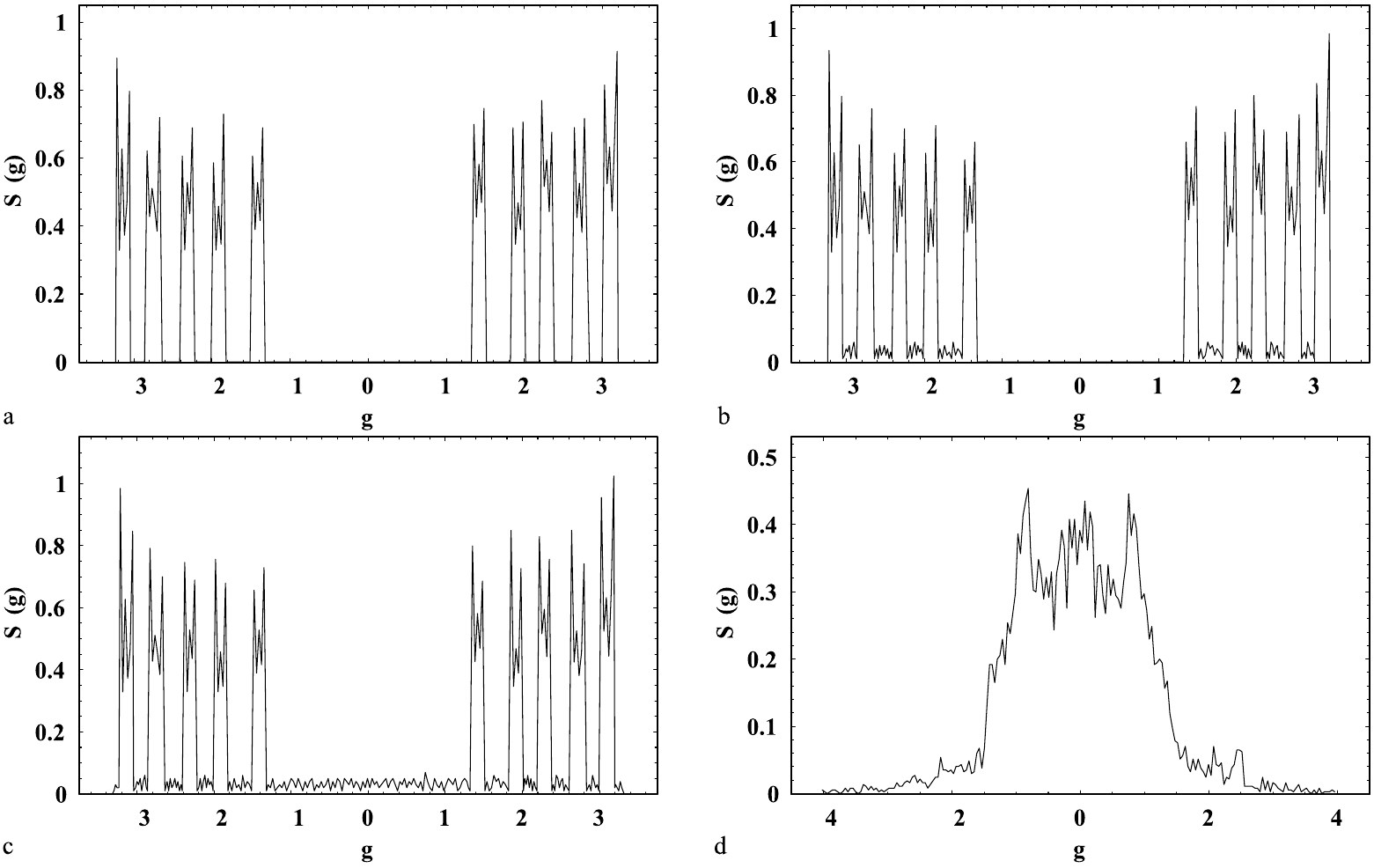}}}
\vskip 0.01cm
\caption{(a-d): Time evolution of the $S(g)$ dynamical spectrum for the sticky orbit. The integration time is (a-upper left): $T = 1150$ time units, (b-upper right): $T = 1670$ time units, (c-lower left): $T = 12200$ time units and (d-lower right): 5 $\times$ $10^4$ time units.}
\end{figure*}
\begin{figure*}[!tH]
\centering
\resizebox{\hsize}{!}{\rotatebox{0}{\includegraphics*{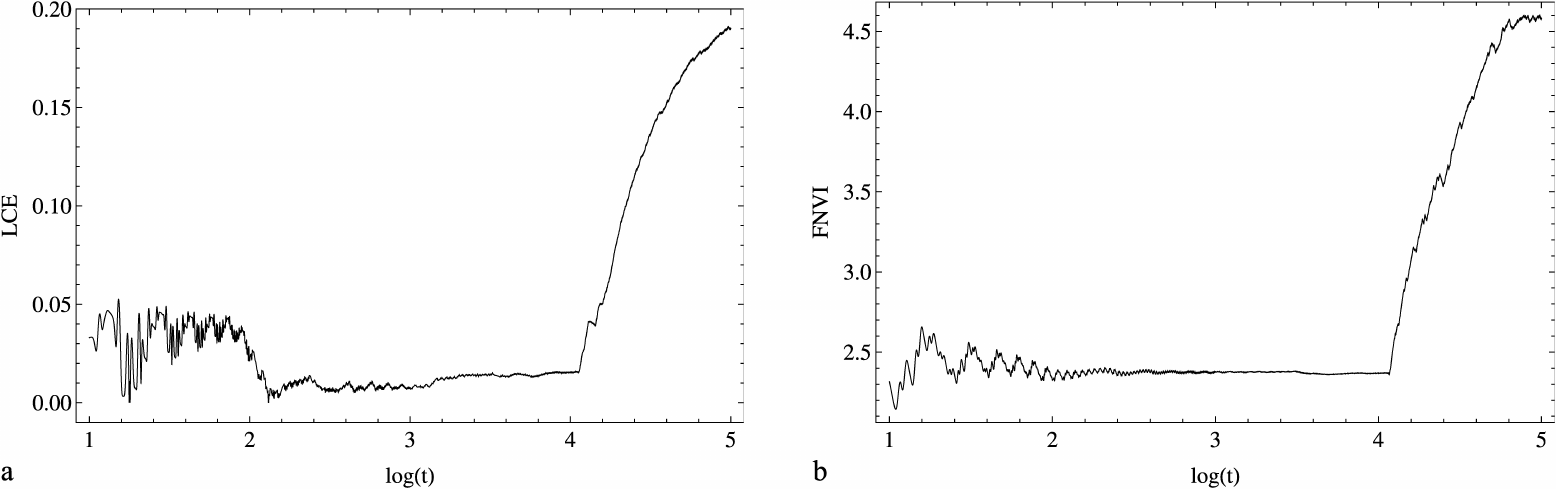}}}
\vskip 0.01cm
\caption{(a-b): Time evolution of the (a-left): the LCE and (b-right): the FNVI for the same sticky orbit.}
\end{figure*}

Finally, in order to have an estimation of the degree of chaos of the 3D dynamical system from another point of view, we have plotted the average value of the maximal LCE versus the value of the energy $h_3$. The results are shown in Figure 12a. Note that the $\langle$ LCE $\rangle$ increases linearly with $h_3$, when the initial value of $z_0$ is 0.01. As we have regular regions and only one unified chaotic sea in the $(x, p_x)$ planes (see Fig. 10a), we calculate the average value of the LCE by taking $10^3$ orbits with different and random initial conditions $(x_0, p_{x0}, z_0 = 0.01)$, $y_0 = p_{z0} = 0$ and $p_{y0} > 0$ in the chaotic domain for every value of the energy $h_3$. Note, that all calculated LCEs are different on the fifth decimal point in the same chaotic region. For different initial value of the $z_0$, our numerical experiments indicate that the relationship between $\langle$ LCE $\rangle$ and $h_3$ remains linear and similar to that shown in Fig. 12a. Here we should point out, that the $\langle$ LCE $\rangle$ of the 3D system is smaller than the corresponding $\langle$ LCE $\rangle$ of the 2D system shown in Fig. 9a. We follow the same procedure using now the dFNVI method. For the same $10^3$ initial conditions $(x_0, p_{x0}, z_0 = 0.01)$, $y_0 = p_{z0} = 0$ and $p_{y0} > 0$ corresponding to chaotic 3D orbits for each value of the energy $h_3$, we calculate the $\langle$ dFNVI $\rangle$. The results are presented in Figure 12b. Again, the $\langle$ dFNVI $\rangle$ increases linearly with $h_3$. The increase of the $\langle$ dFNVI $\rangle$ indicates that the chaoticity of the dynamical system increases as the value of the energy is amplified. For different initial value of the $z_0$, our numerical calculations indicate that the relationship between $\langle$ dFNVI $\rangle$ and $h_3$ also remains linear and similar to that shown in Fig. 12b. As the behavior of $\langle$ dFNVI $\rangle$ coincides with the behavior of $\langle$ LCE $\rangle$, we prove the reliability of dFNVI method also in the 3D dynamical system. Of particular interest is the fact, that the $\langle$ dFNVI $\rangle$ of the 3D system is also smaller than the corresponding $\langle$ dFNVI $\rangle$ of the 2D system shown in Fig. 9b. This strongly suggests that the chaoticity in the dynamical system of three degrees of freedom(3D) is smaller that in the corresponding dynamical system of two degrees of freedom (2D), when $h_2 = h_3$.

\section{Application in the case of a sticky orbit}

In this section, we shall try to test the FNVI method's ability posing a somehow more difficult task, that is to distinguish between ordered and chaotic motion, in the case of chaotic orbits that seem to be regular for a number of periods. Such orbits, called ``sticky orbits" and they usually grow near the outer regions of an island of stability [18]. Cantori that exist outside that region may restrict the motion in a very thin layer for a long time interval. After that time interval, the orbit can escape to the surrounding chaotic region, through the gaps of the cantorus. Due to this peculiar behavior, a sticky orbit often resembles an ordered one seen in a Poincar\'{e} surface of section, until the orbit escapes to the chaotic domain.

We follow the evolution of a two-dimensional sticky orbit which produces two sets of five small islands of invariant curves near the $x$ axis, using the PSS technique. This sticky orbit has initial conditions: $x_0 = -1.705$, $y_0 = 0$ and $p_{x0} = 0$, while the value of $p_{y0}$ is obtained from the energy integral (2). Our results are presented in Fig. 13 (a-d). Fig. 13a shows the regions formed in the $(x, p_x)$ phase plane by this orbit, for a period of 1200 time units of numerical integration. We see two sets of five separate islands of invariant curves. The outermost black solid line shown in Fig. 13a is the Zero Velocity Curve (ZVC) defined by Eq. (9). When $T = 1200$ time units (Fig. 13a), one can see that the orbit is chaotic merely zooming in the plot and seeing that the points are not smoothly distributed over a well defined one-dimensional curve. In Fig. 13b the same orbit was calculated for 1700 time units. Now, we observe that in each case the five small islands of invariant curves were joined together. By observing Figs. 13a and 13b, it is obvious that from the pictures presented in the PSS no safe conclusion can be reached concerning the regular or chaotic behavior of this orbit. Fig. 13c depicts the same orbit calculated for a time period of 12500 time units. It is evident, that now the test particle has left the sticky region and begins to enter the vast surrounding chaotic region. In Fig. 13d we have calculated the orbit for 5 $\times$ $10^4$ time units. The chaotic nature of the orbit is now completely revealed, since the scattered points in the $(x, p_x)$ phase plane fill the entire available chaotic domain.

A better and more enlightening view for the evolution of the sticky orbit can be seen using the $S(g)$ dynamical spectrum [33,35]. Fig. 14a shows the $S(g)$ spectrum of the sticky orbit for 1150 time units of numerical integration. Here we observe ten separate $U$-type structure with a large number of additional small and large peaks indicating the sticky nature of the orbit. In Fig. 14b we present the $S(g)$ spectrum computed for a time interval of 1670 time units. In this case, the ten structures have joined together producing two separate new structures. Fig. 14c depicts the $S(g)$ spectrum of the same orbit calculated for 12200 time units of numerical integration. One may observe, that the two separate structures shown in Fig. 14b have now joined together producing a single spectrum. The shape of this spectrum has all the characteristics of a chaotic spectrum and thus strongly indicates that the test particle has undoubtedly left from the sticky region and continued its journey to the surrounding chaotic sea. Fig. 14d shows the $S(g)$ spectrum computed for 5 $\times$ $10^4$ time units. We see that the spectrum has the usual shape of a spectrum corresponding to a chaotic orbit. Here we must point out, that the time intervals regarding the gradual time evolution of the sticky orbit obtained by the $S(g)$ spectrum are very close to those derived by the formation of the regions in the $(x, p_x)$ phase plane, shown in Fig. 13 (a-d).

Now let's use some other dynamical indicators in order to follow the evolution of the sticky orbit and also compare and verify our results. Fig. 15 (a-b) shows the time evolution of the LCE and the FNVI for the same sticky orbit discussed and studied earlier. Fig. 15a shows the time evolution of the LCE. It is evident that after 12600 time units the orbit gradually becomes chaotic. In Fig. 15b we present the time evolution of the FNVI for this orbit. We observe, that the pattern of the time evolution of the FNVI is almost identical with the pattern of the LCE shown in Fig. 15a. Furthermore, the FNVI clearly indicates that after 12100 time units the orbit gradually alters its nature to chaotic. Note that in Fig. 15a-b the $t$ axis is in log scale in order to have a more clear view of the time evolution. Therefore, we may conclude that the results obtained by the PSS technique, the $S(g)$ dynamical spectrum, the LCE and also the FNVI method shown in Figs. 13a-d, 14a-d, 15a and 15b respectively, coincide that the sticky orbit becomes chaotic after a time interval of about 12500 time units of numerical integration. Thus, we may conclude that the new FNVI method has the ability to follow the time evolution of a two-dimensional sticky orbit and provide reliable results regarding the time needed for the transition from sticky motion to chaos.

\section{Comparison with other dynamical indicators}

The results presented in the previous sections reveal, that the dFNVI is a simple, efficient and easy to compute tool for the distinguishing between ordered and chaotic motion in Hamiltonian systems. Implementing the dFNVI method is a very easy computational task, as we only have to follow the evolution of the norm of an orbits's vector throughout the integration time interval and compute the maximum and the minimum value of FNVI when $t \in [200, 1000]$. In the case of regular motion the FNVI after a small transient period of fluctuation it settles down to a value and remains almost constant, while the dFNVI is always smaller than 0.05. In particular, our extensive numerical calculations suggest, that when an orbits is regular the dFNVI is much smaller that 0.01. However, we decided to increase the threshold value to 0.05 in order to obtain more secure and reliable results. On the other hand, in the case of chaotic motion the FNVI continues to fluctuate randomly without giving any sign of convergence, while the numerical criterion, that is the dFNVI is always larger than 0.05. It is exactly this different behavior of the FNVI regarding the convergence, that makes it an ideal tool of chaos detection. However, the results obtained by the FNVI are only qualitative. Thus, we established a numerical criterion the dFNVI, which can provide quantitative results about the nature of an orbit. As we have seen in the previous sections, the results obtained by the dFNVI are completely reliable, conclusive and beyond any doubt. The dFNVI helps us decide the chaotic or regular nature of orbits faster and with less computational effort than the estimation of the maximal LCE. This happens because the time needed for the LCE in order to give a clear and undoubted indication of convergence to non-zero values is usually much greater than the time needed for the dFNVI to provide reliable evidence regarding the character of an orbit.

Many other dynamical indicators have been introduced in the recent years, some of which are compared in this section with the dFNVI method. In order to verify and prove, once more, the effectiveness and the reliability of our new method, we shall compare our results with four other well known dynamical indicators: (a) The Fast Lyapunov Indicator (FLI), (b) The Relative Lyapunov Indicator (RLI), (c) The Smaller Alignment Index (SALI) and (d) The Generalized Alignment Index (GALI). At this point we believe, that it should be wise to recall the definitions of these indicators, in order to help the reader to compare the results of each method.

The FLI introduced in [8, 9] has been proved a very fast, reliable and effective tool, which can be defined as
\begin{equation}
\rm FLI(t) = \log \| \vec{w}(t) \|, \ \ \ \ t \leq t_{max},
\end{equation}
where $\vec{w}(t)$ is a deviation vector. The computation of the FLI on a relatively short time $t_{max}$ is enough to discriminate between chaotic and regular orbits. The FLI of a regular orbit increases linearly, while for a chaotic orbit, the FLI increases super-exponentially. Moreover, the FLI may be used to discriminate among regular motion between non resonant and resonant orbits. In order to avoid the linear trend of the FLI along regular orbits, we can use the de-trended FLI (DFLI), which is simply the FLI divided by $t$. Thus,
\begin{equation}
\rm DFLI(t) = \log \left(\frac{\| \vec{w}(t) \|}{t}\right), \ \ \ \ t \leq t_{max}.
\end{equation}
In this case, the classification of an orbit is based on the clearly distinct behavior of the DFLI pattern. In particular, in the case of a regular orbit the DFLI is almost constant with values lower than 10, while in the case of a chaotic orbit, we observe a very rapid increase of the DFLI. We believe, that using definition (13) instead of (12) we can obtain better and more conclusive results. The DFLI is a relatively fast indicator, as it needs only about 5 $\times$ $10^3$ times units in order to provide reliable results regarding the nature of an orbit. Note, that for the calculation of the DFLI, we need the assistance of the variational equations and of course the computation of the deviation vector $\vec{w}(t)$.

Another interesting dynamical indicator is the RLI proposed by S\'{a}ndor et al [23]. The RLI is practically the absolute value of the difference of $L_t$ of two initially nearby orbits and can be defined as
\begin{equation}
\rm RLI(t) = \left| L \left(\vec{x} + \delta \vec{x}, t \right) - L \left(\vec{x}, t \right) \right|,
\end{equation}
where
\begin{equation}
L\left(\vec{x}, t\right) = \frac{1}{t} \ln \frac{\|\vec{w_1}(t)\|}{\|\vec{w_1}(0)\|}, \ \ \
L\left(\vec{x} + \delta \vec{x}, t\right) = \frac{1}{t} \ln \frac{\|\vec{w_2}(t)\|}{\|\vec{w_2}(0)\|},
\end{equation}
while $\left| \delta \vec{x} \right| \ll 1$. Very small values of the RLI $\left(\rm RLI < 10^{-11} \right)$  denote ordered motion, while larger values $\left(\rm RLI > 10^{-4} \right)$ denote chaotic behavior (for more information on the RLI method see [23]). The RLI method needs at least about 5 $\times$ $10^4$ times units in order to provide reliable results regarding the nature of an orbit. We must point out, that the computation of the RLI requires the time evolution of two orbits and two deviation vectors ($\vec{w}_1(t)$ and $\vec{w}_2(t)$ one for each orbit), while the computation of the dFNVI is much faster as we compute only the evolution of the main orbit without the use of any deviation vector.

Using the variational equations and the evolution of the deviation vectors we can compute the SALI [25, 26]. The basic idea behind the SALI method is the introduction of a simple quantity that clearly indicates if a deviation vector is aligned with the direction of the eigenvector which corresponds to the maximal LCE. In general, any two randomly chosen initial deviation vectors $\vec{w}_1(0)$ and $\vec{w}_2(0)$ will become aligned with the most unstable direction and the angle between them will rapidly tend to zero [2]. Thus, we check if the two vectors have the same direction in phase space, which is equivalent to the computation of the above-mentioned angle. More specifically, we follow simultaneously the time evolution of an orbit with initial condition $\vec{x}(0)$ and two deviation vectors with initial conditions $\vec{w}_1(0)$ and $\vec{w}_2(0)$. As we are only interested in the directions of these two vectors we normalize them, at every time step, keeping their norm equal to 1. This controls the exponential increase of the norm of the vectors and avoids overflow problems. Since, in the case of chaotic orbits the normalized vectors point to the same direction and become equal or opposite in sign, the minimum of the norms of their sum (\textit{antiparallel alignment index}) or difference (\textit{parallel alignment index}) tends to zero. So, the SALI is defined as
\begin{equation}
\rm SALI(t) = min(d_-(t), d_+(t)),
\end{equation}
where $d_-(t) = \left \| \frac{\vec{w}_1(t)}{\| \vec{w}_1(t) \|} - \frac{\vec{w}_2(t)}{\| \vec{w}_2(t) \|} \right \|$ and $d_+(t) = \left \| \frac{\vec{w}_1(t)}{\| \vec{w}_1(t) \|} + \frac{\vec{w}_2(t)}{\| \vec{w}_2(t) \|} \right \|$ are the parallel and the antiparallel alignment indices respectively. From the above definition it is evident that SALI(t) $\in [0, \sqrt{2}]$ and when SALI = 0 the two normalized vectors have the same direction, being equal or opposite. Implementing the SALI method is an easy computational task, as we only have to follow the evolution of an orbit and of two deviation vectors, computing in every time step the minimum norm of the difference and the addition of these vectors. In the case of chaotic motion the SALI eventually tends exponentially to zero, reaching rapidly very small values or even the limit of the accuracy of the computer $(10^{-16})$. On the other hand, in the case of ordered motion the SALI fluctuates around non-zero values. The SALI has a clear physical meaning as zero, or a very small value of the index, signifies the alignment of the two deviation vectors. An advantage of the method is that the index ranges in a defined interval (SALI $\in [0, \sqrt{2}])$ and so very small values of the SALI (e.g., smaller than $10^{-8}$) establish the chaotic nature of an orbit beyond any doubt. The SALI is a very fast indicator, as it needs only about $10^3$ times units in order to reveal the regular or chaotic character of an orbit.

Recently, Skokos et al [27] generalized and improved considerably the SALI method mentioned above, by introducing the GALI method. This indicator retains the advantages of the SALI – i.e. its simplicity and efficiency in distinguishing between regular and chaotic motion – but, in addition, is faster than the SALI, displays power law decays that depend on torus dimensionality and can also be applied successfully to cases where the SALI is inconclusive, like in the case of chaotic orbits whose two largest Lyapunov exponents are equal or almost equal. For the computation of the GALI we use information from the evolution of more than two deviation vectors with respect to the reference orbit, while SALI's computation requires the evolution of only two such vectors. In particular, $\rm GALI_k$ is proportional to ``volume" elements formed by $k$ initially linearly independent unit deviation vectors whose magnitude is normalized to unity at every time step. If the orbit is chaotic, $\rm GALI_k$ goes to zero exponentially fast by the law
\begin{equation}
\rm GALI_k(t) \propto e ^{-\left[(\sigma_1 - \sigma_2) + (\sigma_1 - \sigma_3) + ... + (\sigma_1 - \sigma_k)\right] t}.
\end{equation}
If, on the other hand, the orbit lies in an $N$-dimensional torus, $\rm GALI_k$ displays the following behaviors: Either
\begin{equation}
\rm GALI_k(t) \approx constant  \ \ \ \ for \ \ 2 \leq k \leq N,
\end{equation}
or, if $N < k \leq 2N$, it decays with different power laws, depending on the number $m$ of deviation vectors which initially lie in the tangent space of the torus, i.e.,
\begin{equation}
\rm  GALI_k(t) \propto \left\{
\begin{array}{l l}
\frac{1}{{{t}^{2\left(k - N \right) - m}}} & \quad \textrm{if \ \ $N < k \leq 2N$ \ \ and \ \ $0 \leq m < k - N$} \\
 \\
\frac{1}{{{t}^{k - N}}} & \quad \textrm{if \ \ $N < k \leq 2N$ \ \ and \ \ $m \geq k - N$.} \\
\end{array} \right.
\end{equation}
So, the GALI method allows us to study more efficiently the geometrical properties of the dynamics in the neighborhood of an orbit, especially in higher dimensions, where it allows for a much faster determination of its chaotic nature, overcoming the limitations of the SALI method. In the case of regular motion, $\rm GALI_k$ is either a constant, or decays by power laws that depend on the dimensionality of the subspace in which the orbit lies, which can prove useful e.g., if our orbits are in a ``sticky" region, or if our system happens to possess fewer or more than $N$ independent integrals of the motion (i.e. is partially integrable or super-integrable respectively). It should be mentioned, that both the SALI and the GALI methods are very reliable and efficient. However, in order to apply them we have to use the variational equations and also several sets of deviation vectors. On the contrary, the dFNVI method is much faster than these methods because it does nor require the use of the variational equations and the deviation vectors. Simply we follow the time evolution of an orbit with initial condition $\vec{x}(0)$ and computing its norm by integrating only the basic equations of motion.
\begin{figure*}
\centering
\resizebox{\hsize}{!}{\rotatebox{0}{\includegraphics*{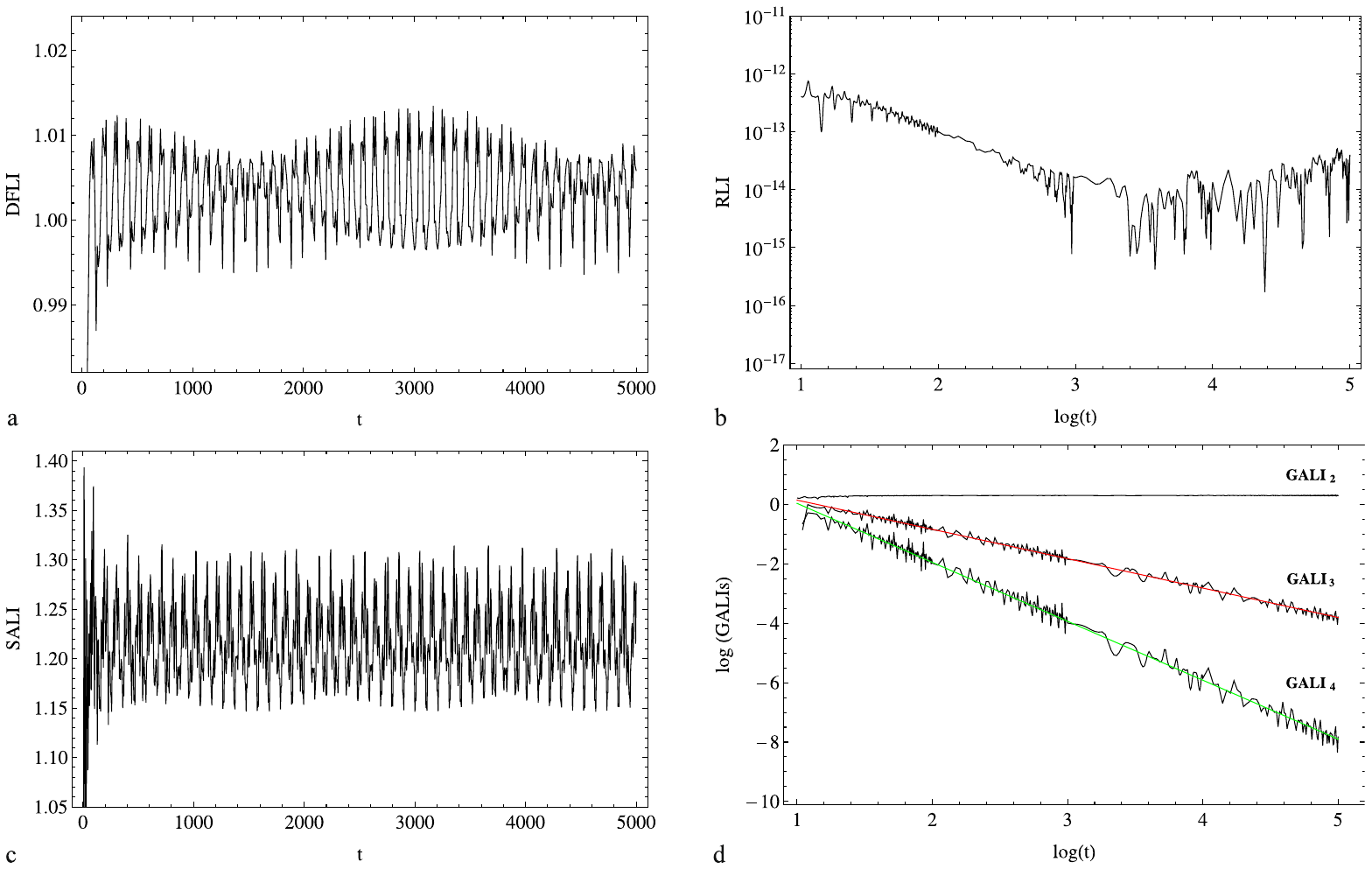}}}
\vskip 0.01cm
\caption{(a-d): Evolution of different dynamical indicators for a regular 2D orbit of the dynamical system (7) (a-upper left): DFLI, (b-upper right): RLI, (c-lower left): SALI and (d-lower right): $\rm GALI_k$. The initial conditions of the 2D orbit and more details are provided in the text.}
\end{figure*}
\begin{figure*}
\centering
\resizebox{\hsize}{!}{\rotatebox{0}{\includegraphics*{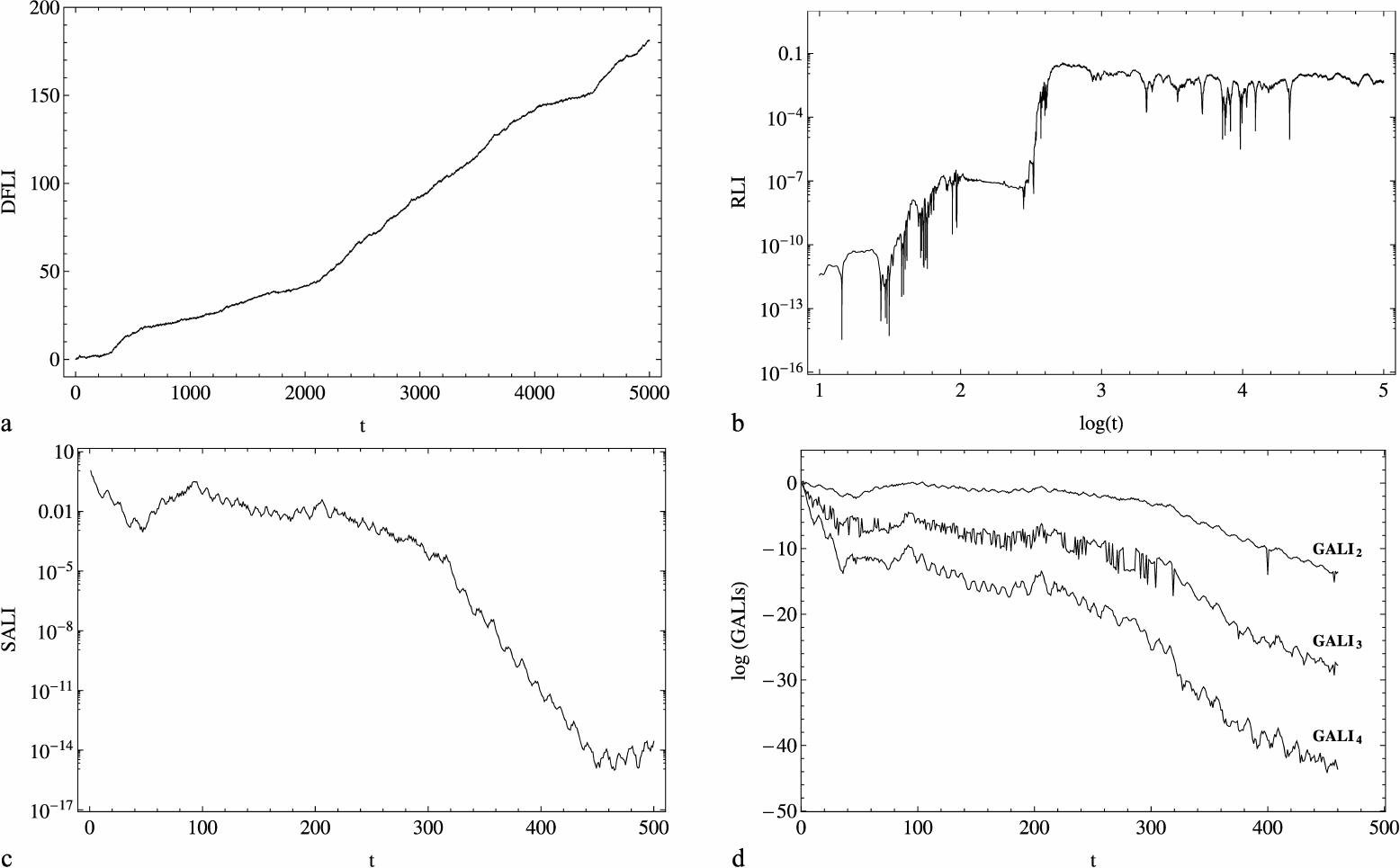}}}
\vskip 0.01cm
\caption{(a-d): Similar to Fig. 16 a-d but for a chaotic 2D orbit of the dynamical system (7). The initial conditions of the 2D orbit and more details are provided in the text.}
\end{figure*}
\begin{figure*}
\centering
\resizebox{\hsize}{!}{\rotatebox{0}{\includegraphics*{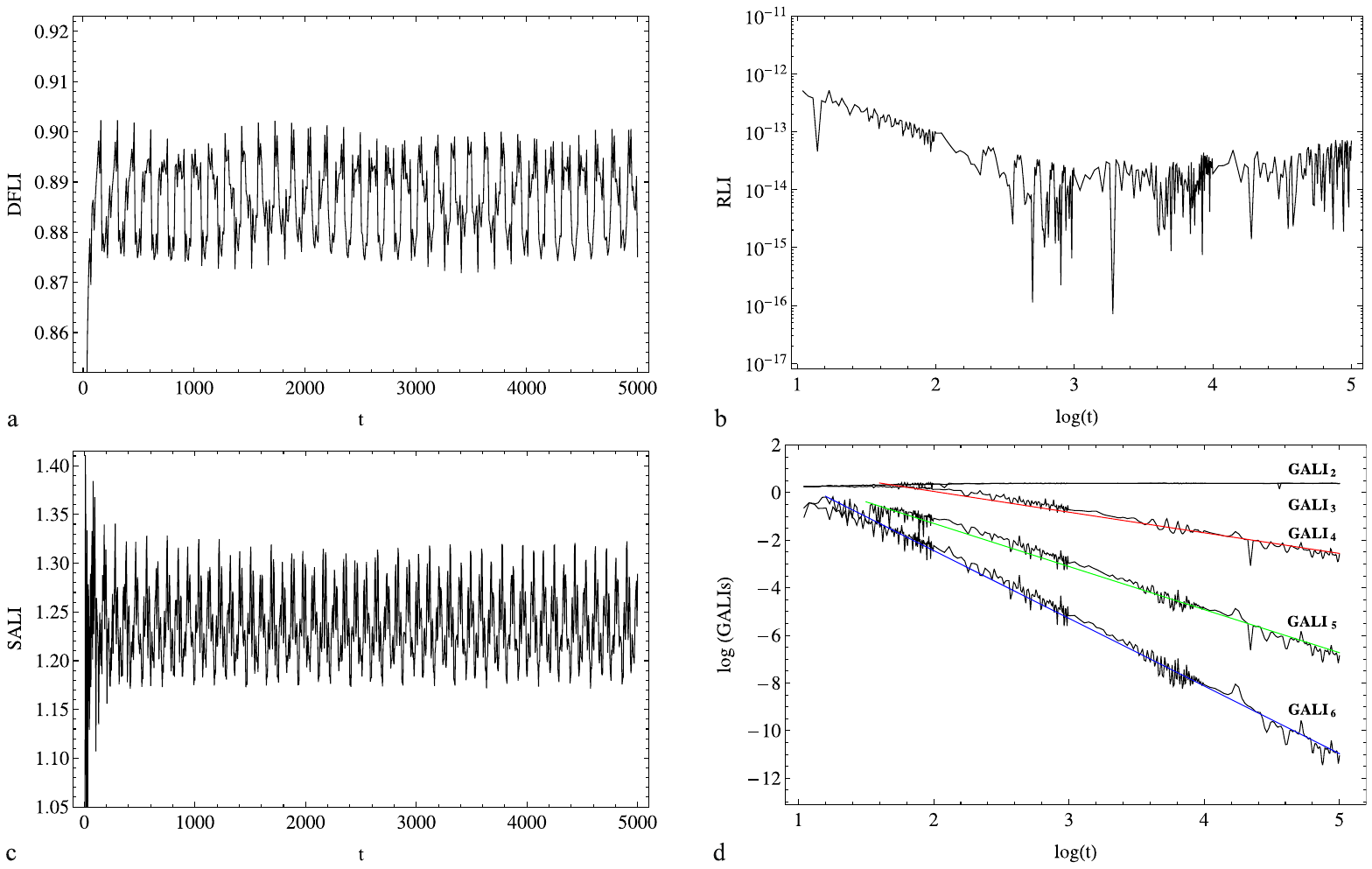}}}
\vskip 0.01cm
\caption{(a-d): Evolution of different dynamical indicators for a regular 3D orbit of the dynamical system (6) (a-upper left): DFLI, (b-upper right): RLI, (c-lower left): SALI and (d-lower right): $\rm GALI_k$. The initial conditions of the 3D orbit and more details are provided in the text.}
\end{figure*}
\begin{figure*}
\centering
\resizebox{\hsize}{!}{\rotatebox{0}{\includegraphics*{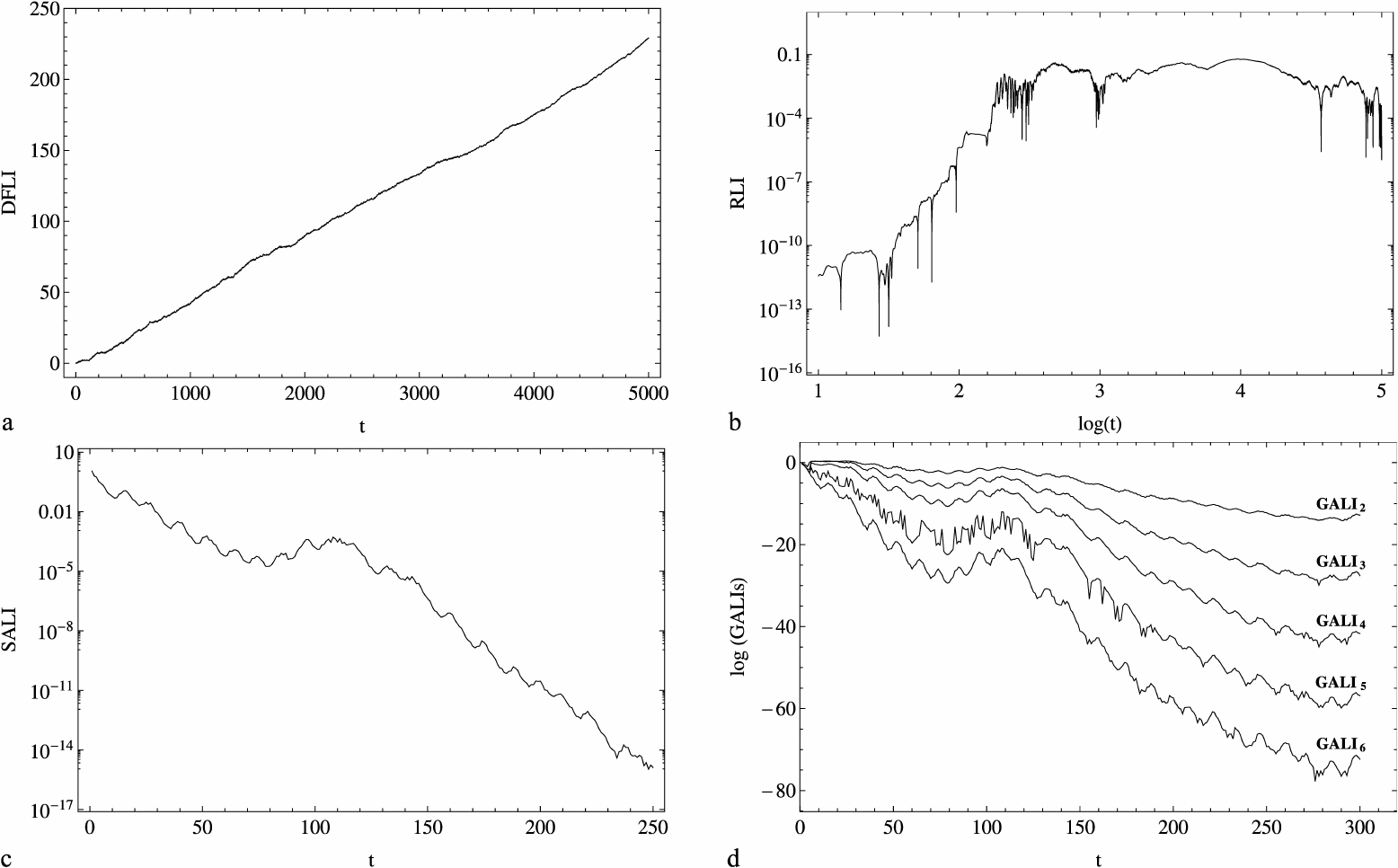}}}
\vskip 0.01cm
\caption{(a-d): Similar to Fig. 18 a-d but for a chaotic 3D orbit of the dynamical system (6). The initial conditions of the 3D orbit and more details are provided in the text.}
\end{figure*}

In what follows, we shall present and compare representative examples for 2D and 3D orbits of the dynamical systems (7) and (6) respectively, using the above four indicators. In Figure 16a-d, we can see the results provided by the four indicators for a regular 2D orbit with initial conditions: $x_0 = 0.27$, $y_0 = 0$ and $p_{x0} = 0$, while the initial value of $p_{y0}$ is found from the energy integral (7). The value of the energy is $h_2 = 1$. In Fig. 16a we present the evolution of the DFLI for a time period of 5 $\times$ $10^3$ time units. We observe that the DFLI fluctuates around small values (DFLI $<$ 10), which indicates regular motion. The initial deviation vector, $\vec{w} = (dx, dy, dp_x, dp_y)$, used for the orbit shown in Fig. 16a is $\vec{w}(0) = (1, 0, 0, 0)$. Fig. 16b depicts the evolution of the RLI for a time interval of $10^5$ time units, for the same regular orbit. The RLI clearly reveals the ordered nature of the orbit, since RLI $<$ $10^{-12}$ throughout the time evolution. The initial deviation vectors, used for the computation of the RLI are $\vec{w}_1(0) = (1, 0, 0, 0)$ and $\vec{w}_2(0) = (1, 0, 0, 0)$, while $\delta x = 10^{-12}$. Note that the $t$ axis is in log scale. In Fig. 16c we present the evolution of SALI for a time interval of 5 $\times$ $10^3$ time units. Also this method indicates the regular character of the orbit, since it exhibits small fluctuations around non-zero values. The initial deviation vectors used for the orbit of Fig. 16c are $\vec{w}_1(0) = (1, 0, 0, 0)$ and $\vec{w}_2(0) = (0, 0, 1, 0)$, but in general, any other choice of initial values leads to similar behavior of the SALI. From Eqs. (18) and (19) it follows that in the case of a Hamiltonian system with $N$ = 2 degrees of freedom $\rm GALI_2$ will always remains different from zero, while $\rm GALI_3$ and $\rm GALI_4$ should decay to zero following a power law, whose exponent depends on the number $m$ of the deviation vectors that are initially tangent to the torus on which the orbit lies. Now, for the regular orbit of the 2D Hamiltonian (7) and for a random choice of initial deviation vectors, we expect the $\rm GALI_k$ indices to behave as
\begin{eqnarray}
\rm GALI_2&(t)& \propto constant, \nonumber \\
\rm GALI_3&(t)& \propto \frac{1}{t^2}, \nonumber \\
\rm GALI_4&(t)& \propto \frac{1}{t^4}.
\end{eqnarray}
Fig. 16d shows the behavior of $\rm GALI_k$ for the same regular 2D orbit. Indeed, the $\rm GALI_k$ indices obey to the power laws respectively given in Eq. (20). In Fig. 16d with red color we plot the power law $1/t^{2}$, while the power law $1/t^{4}$ is colored in green. Note that the $t$ axis is in log scale. The evolution of the FNVI for this regular 2D orbit, for a time interval of $10^5$ time units, is presented in Fig. 3a, while dFNVI = 0.009. We see, that all the outcomes derived using five different dynamical indicators coincide to the ordered nature of the orbit.

Figure 17a-d is similar to Fig. 16a-d but for a chaotic 2D orbit. In Fig. 17a-d, we present the results provided by the four dynamical indicators for a chaotic 2D orbit with initial conditions: $x_0 = 0.82$, $y_0 = 0$ and $p_{x0} = 0$, while the initial value of $p_{y0}$ is found from the energy integral (7). The value of the energy is $h_2 = 1$. In Fig. 17a we present the evolution of the DFLI for a time period of 5 $\times$ $10^3$ time units. We observe that the DFLI displays an exponential growth with time, which indicates chaotic motion. The initial deviation vector used for the orbit shown in Fig. 17a is $\vec{w}(0) = (1, 0, 0, 0)$. Fig. 17b depicts the evolution of the RLI for a time interval of $10^5$ time units, for the same chaotic orbit. Approximately for the first 350 time units, the RLI has low values (RLI $\lesssim 10^{-7}$) but for the rest time interval RLI $\gtrsim 10^{-4}$. Note that the $t$ axis is in log scale. Therefore, we may conclude that the orbit is chaotic. Note, that the RLI method gives inconclusive or even misleading results for very small time intervals of the numerical integration. The initial deviation vectors, used for the computation of the RLI are $\vec{w}_1(0) = (1, 0, 0, 0)$ and $\vec{w}_2(0) = (1, 0, 0, 0)$, while $\delta x = 10^{-12}$. In Fig. 17c we present the evolution of the SALI. We see that after a small transient time, the SALI falls abruptly to zero. At $t \approx 450$ time units the SALI becomes zero, as it has reached the limit of the accuracy of the computer $(10^{-16})$, which means that the two deviation vectors have the same direction. Thus, after $t \approx 450$ time units the two normalized vectors are represented by exactly the same numbers in the computer and we can safely argue, that to this accuracy the orbit is chaotic. The initial deviation vectors used for the orbit of Fig. 17c are $\vec{w}_1(0) = (1, 0, 0, 0)$ and $\vec{w}_2(0) = (0, 0, 1, 0)$. From Eq. (17) it is evident that in the case of a Hamiltonian system with $N$ = 2 degrees of freedom, $\rm GALI_k$ indices should decrease exponentially approaching the zero-value. In Fig. 17d we observe the behavior of $\rm GALI_k$ for the same chaotic 2D orbit. Indeed, the $\rm GALI_k$ indices obey to the exponential law given in Eq. (17). The evolution of the FNVI for this chaotic 2D orbit, for a time interval of $10^5$ time units, is presented in Fig. 3b, while dFNVI = 0.45. We see, that all the outcomes obtained using five different dynamical indicators coincide to the chaotic nature of the orbit.

We shall now proceed, presenting and comparing two more examples regarding 3D orbits of the Hamiltonian system (6). In Figure 18a-d, we can see the results provided by the four indicators for a regular 3D orbit with initial conditions: $x_0 = 0.09$, $y_0 = 0, z_0 = 0.1, p_{x0} = p_{z0} = 0$, while the initial value of $p_{y0}$ is found from the energy integral (6). The value of the energy is $h_3 = 1$. In Fig. 18a we present the evolution of the DFLI for a time period of 5 $\times$ $10^3$ time units. We observe that the DFLI fluctuates around small values (DFLI $<$ 10), which indicates regular motion. The initial deviation vector, $\vec{w} = (dx, dy, dz, dp_x, dp_y, dp_z)$, used for the orbit shown in Fig. 18a is $\vec{w}(0) = (1, 0, 0, 0, 0, 0)$. Fig. 18b depicts the evolution of the RLI for a time interval of $10^5$ time units, for the same regular orbit. The RLI clearly reveals the ordered character of the orbit, since RLI $<$ $10^{-12}$ throughout the time evolution. Note that the $t$ axis is in log scale. The initial deviation vectors, used for the computation of the RLI are $\vec{w}_1(0) = (1, 0, 0, 0, 0, 0)$ and $\vec{w}_2(0) = (1, 0, 0, 0, 0, 0)$, while $\delta x = 10^{-12}$. In Fig. 18c we present the evolution of SALI for a time interval of 5 $\times$ $10^3$ time units. Also this method indicates the regular character of the orbit, since it exhibits small fluctuations around non-zero values (SALI $\in [1, \sqrt{2}]$). The initial deviation vectors used for the orbit of Fig. 18c are $\vec{w}_1(0) = (1, 0, 0, 0, 0, 0)$ and $\vec{w}_2(0) = (0, 0, 0, 1, 0, 0)$, but in general any other choice of initial values leads to similar behavior of the SALI. From Eqs. (18) and (19) it follows that in the case of a Hamiltonian system with $N$ = 3 degrees of freedom $\rm GALI_2$ and $\rm GALI_3$ will always remain different from zero, while $\rm GALI_4$, $\rm GALI_5$ and $\rm GALI_6$ should decay to zero following a power law, whose exponent depends on the number $m$ of deviation vectors that are initially tangent to the torus on which the orbit lies. Now, for the regular orbit of the 3D Hamiltonian (6) and for a random choice of initial deviation vectors, we expect the $\rm GALI_k$ indices to behave as
\begin{eqnarray}
\rm GALI_2&(t)& \propto constant, \nonumber \\
\rm GALI_3&(t)& \propto constant, \nonumber \\
\rm GALI_4&(t)& \propto \frac{1}{t^2}, \nonumber \\
\rm GALI_5&(t)& \propto \frac{1}{t^4}, \nonumber \\
\rm GALI_6&(t)& \propto \frac{1}{t^6}.
\end{eqnarray}
Fig. 18d shows the behavior of $\rm GALI_k$ for the same regular 3D orbit. Indeed, the $\rm GALI_k$ indices obey to the power laws respectively given in Eq. (21). In Fig. 18d with red color we plot the power law $1/t^{2}$, the power law $1/t^{4}$ is colored in green, while the power law $1/t^{6}$ is plotted with blue color. Note that the $t$ axis is in log scale. With a more closer look in Fig. 18d, we observe that there is a superposition of $\rm GALI_2$ and $\rm GALI_3$ as they evolve almost identically. The evolution of the FNVI for this regular 3D orbit, for a time interval of $10^5$ time units, is presented in Fig. 4a, while dFNVI = 0.006. We see, that once more, all the outcomes derived using five different dynamical indicators coincide to the ordered nature of the orbit.

Finally, Figure 19a-d is similar to Fig. 18a-d but for a chaotic 3D orbit. In Fig. 19a-d, we present the results provided by the four dynamical indicators for a chaotic 3D orbit with initial conditions: $x_0 = 0.86, y_0 = 0, z_0 = 0.1, p_{x0} = p_{z0} = 0$, while the initial value of $p_{y0}$ is found from the energy integral (6). The value of the energy is $h_3 = 1$. In Fig. 19a we present the evolution of the DFLI for a time period of 5 $\times$ $10^3$ time units. We observe that the DFLI displays a very rapid growth with time, which indicates chaotic motion. The initial deviation vector used for the orbit shown in Fig. 19a is $\vec{w}(0) = (1, 0, 0, 0)$. Fig. 19b depicts the evolution of the RLI for a time interval of $10^5$ time units, for the same chaotic orbit. Approximately for the first 200 time units, the RLI has low values (RLI $\lesssim 10^{-6}$) but for the rest time interval RLI $\gtrsim 10^{-4}$. Note that the $t$ axis is in log scale. Therefore, we may conclude that the orbit is chaotic. Note, that again the RLI method gives inconclusive or even misleading results for very small time intervals of the numerical integration. The initial deviation vectors, used for the computation of the RLI are $\vec{w}_1(0) = (1, 0, 0, 0, 0, 0)$ and $\vec{w}_2(0) = (1, 0, 0, 0, 0, 0)$, while $\delta x = 10^{-12}$. In Fig. 19c we present the evolution of the SALI. We see that the SALI decreases very rapidly approaching the zero-value. At $t \approx 250$ time units the SALI becomes zero, as it has reached the limit of the accuracy of the computer $(10^{-16})$, which means that the two deviation vectors have the same direction. Thus, after $t \approx 250$ time units the two normalized vectors are represented by exactly the same numbers in the computer and we can safely argue, that to this accuracy the orbit is chaotic. The initial deviation vectors used for the orbit of Fig. 19c are $\vec{w}_1(0) = (1, 0, 0, 0, 0, 0)$ and $\vec{w}_2(0) = (0, 0, 0, 1, 0, 0)$. From Eq. (17) it is evident that in the case of a Hamiltonian system with $N$ = 3 degrees of freedom, all $\rm GALI_k$ indices should decrease exponentially approaching the zero-value. In Fig. 19d we observe the behavior of $\rm GALI_k$ for the same chaotic 3D orbit. Indeed, all the $\rm GALI_k$ indices obey to the theoretically predicted law given in Eq. (17). The evolution of the FNVI for this chaotic 3D orbit, for a time interval of $10^5$ time units, is presented in Fig. 4b, while dFNVI = 0.52. We see, that all the outcomes obtained using five different dynamical indicators coincide to the chaotic nature of the orbit.

From the above examples, it becomes evident that all the different methods we used (DFLI, RLI, SALI and GALI) need not only the computation of the basic equations of motion but also the simultaneous computation of the so-called variational equations and of course several sets of deviation vectors. As the total number of the equations and inevitably the total number of the variables increases, we have also an increase to the time needed from the computational code in order to integrate the system of the differential equations and provide the results. Thus, our proposed FNVI method has one important advantage over all the other methods mentioned above, since it requires only the computation of the basic set of the equations of motion without any use of variational equations and deviation vectors. Let us present a specific illuminating example regarding the speed of the methods. Figure 20 presents the CPU time in h for each dynamical method required, using the same technique described in section 3 and of course the same processor, in order to construct the grid-plot shown in Fig. 7a. We observe, that the dFNVI method, needs only about 7 h of CPU time. The total time required by the $S(g)$ spectrum is similar to the time needed for the dFNVI method. On the other hand, if we choose to use methods such as the DFLI, SALI and GALI$_k$, that need the additional computation of the variational equations, the total time required for the completion of the computational task increases significantly. Furthermore, we see from Fig. 20 that the most time consuming methods are the LCE and RLI. This is true, because when we use these two methods, we have to integrate each orbit at least for $10^4$ time units in order to obtain reliable information about the ordered or chaotic nature of the orbit. Thus, it becomes more than obvious, than the use of these two methods doubles the total CPU time needed by the dFNVI method. Therefore, our new method is not only reliable but also very fast compared with other methods and can be applied easily when we need to integrate a large number of orbits so as to construct a grid-plot of initial conditions (see Figs. 7 and 10).
\begin{figure}[!tH]
\centering
\resizebox{\hsize}{!}{\rotatebox{0}{\includegraphics*{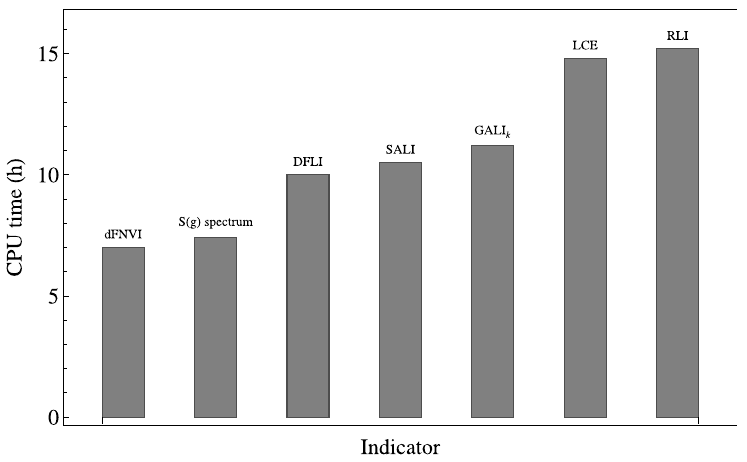}}}
\caption{The total CPU time in h for each dynamical method required, in order to construct the grid-plot shown in Fig. 7a. See text for more details.}
\end{figure}

Apart from the dynamical indicators described and applied previously, there are also other methods for distinguishing between ordered and chaotic motion, such as the mean exponential growth of nearby orbits (MEGNO) [4, 5], the power spectra of deviation vectors [30], the method of the low frequency power (LFP) [13, 29], the ``0--1" test [10], as well as some other more recently introduced techniques [11, 24]. Moreover, a method that have also the ability to identify sticky orbits is the FtDt method [12]. This method uses the Fast Fourier Transform of a series of time interval, each one representing the time that elapsed between two successive points on the Poincar\'{e} surface of section. However, we do not feel that it is necessary to provide examples using all of these methods. We believe, that the comparison between the FNVI method and four well-known dynamical indicators proved its reliability and also revealed its remarkable efficiency.

\section{Discussion and conclusions}

In the present paper, we have tried to introduce a new, fast, efficient and easy to compute method in order to distinguish between order and chaos in 2D and 3D autonomous Hamiltonian systems. We have also conducted a detailed study of the behavior of this new indicator for both chaotic and regular orbits. The main results of this research can be summarized as follows:

\textbf{1.} The FNVI method proves to be an ideal detector of chaoticity independent of the dimensions of the dynamical system. It displays large, abrupt and random fluctuations for chaotic orbits, while it remains almost constant for ordered ones and so it clearly distinguishes between these two cases. Its main advantages are its simplicity, efficiency and reliability as it can rapidly and accurately determine the chaotic versus ordered nature of a given orbit. Of course, the FNVI method can provide only qualitative results regarding the nature of an orbit and therefore, we have to inspect the shape of FNVI each time in order to characterize an orbit. Obviously, this is not very practical when someone wants to check a large volume of orbits, so as to form an idea about the global structure of the dynamical system. Therefore, we proceeded one step further establishing a numerical criterion in order to quantify the results obtained by the FNVI method. This criterion derived by exploiting the shape of the FNVI. When the orbit is regular the FNVI remains almost constant, while in the case of a chaotic orbit it displays high fluctuations. Thus, we calculate the maximum and the minimum value of FNVI when $t \in [200, 1000]$. We choose this particular time interval because in the case where the orbit is regular, for $t \lesssim 200$ time units there is a transient period of fluctuation, which may cause a problem to our criterion. Using the above procedure we defined the dFNVI. The value of the dFNVI can provide us the quantitative criterion that we seek. We point out, that it is not very easy to define a threshold value, so that the dFNVI being larger than this value reliably signifies chaoticity. Nevertheless, extensive numerical experiments in both dynamical systems indicate that in general, a good guess for this value could be 0.05. Thus, when dFNVI $\geqslant$ 0.05 the orbit is chaotic, while when dFNVI $<$ 0.05 the orbit ir ordered. This threshold value applies in both 2D and 3D dynamical systems.

\textbf{2.} We emphasize, that the main advantage of the FNVI method, is that it needs only the computation of the basic equations of motion. In particular, we only have to follow the evolution of the norm of the vector $\vec{x}(t)$ and compare its value in every time step throughout the entire time interval of the numerical integration with the initial value $\vec{x}(0)$ according to Eq. (4). The FNVI method is very fast, as it requires only $10^3$ time units of numerical integration, in order to provide reliable and conclusive results regarding the character of an orbit. Of course, there are also other methods and indicators capable to discriminate between ordered and chaotic orbits. However, the vast majority of them need the computation of the variational equations and inevitably the use of several sets of initial deviation vectors in order to function and provide their results. The additional use and therefore the computation of more equations increases significantly the total time needed from the computational routine to provide the output in each case (see Fig. 20). Thus, our proposed FNVI method has an important advantage over most the other dynamical methods, since it requires only the computation of the basic set of the equations of motion. Therefore, our method is very fast which makes it is an ideal tool when we need to calculate a large number of orbits so as to construct a grid of initial conditions.

\textbf{3.} Using the numerical quantified version of FNVI, that is the dFNVI, we are in a position to characterize reliable an orbit of being chaotic or ordered. Exploiting the advantages of the dFNVI method, we have constructed detailed phase-space portraits (grid-plots) both for 2D and 3D Hamiltonian systems, where the chaotic and ordered regions are clearly distinguished (see Figs. 7 and 10). We were thus able to trace in a fast and systematic way very small islands of ordered motion, whose detection by traditional methods would be very difficult, if not impossible, and moreover very time consuming. This approach is therefore, expected to provide useful tools for the location of stable periodic orbits, or the computation of the phase-space volume occupied by ordered or chaotic motion in multi-dimensional systems $(N \geq 3$), where the PSS plots can not be easily visualized and interpreted and furthermore, very few other similar techniques of practical value are available. Here, we have to point out that grid-plots can be also constructed using other dynamical indicators. For the reasons mentioned in the previous point, the dFNVI method is much more faster than most of the other methods.

\textbf{4.} We used the new FNVI method in order to follow the time evolution of a sticky orbit. The study of the evolution of sticky orbits (orbits that, although they are chaotic, they are restricted in a thin layer of phase space for a large value of integration time) is crucial for the understanding of the structure of the dynamical system. The ability of our new method to extract reliable information concerning the chaotic or regular behavior of an orbit enables us to overcome the ambiguity in the case of sticky orbits. In the case of sticky orbits, the distinction can not be easily made using the PSS technique, as a sticky orbit can be mistaken for an ordered one, if not enough time is given and the orbit has not yet escaped to the surrounding chaotic domain (see Fig. 13a-d). Moreover, the LCE is not sensitive enough to identify the difference between a regular and a sticky orbit (see Fig. 15a). On the contrary, the FNVI method can be applied to obtain a clear and reliable distinction between ordered and sticky orbits (see Fig. 15b) and this could be regarded as a major advantage of our method.

In the present research, we have tested approximately $10^6$ orbits (2D and 3D) in the dynamical systems (6) and (7). Since our results are consistent, we may safely conclude that the FNVI method (and also its numerical version the dFNVI) is a very fast and reliable tool for distinguishing between order and chaos in Hamiltonian systems. It is our future plans, to apply this new method in other kinds of interesting potentials and also in more complicated dynamical systems, in order to test further its efficiency and reliability. Furthermore, we will apply this new method in order to study the evolution of three-dimensional sticky orbits. Moreover, with additional theoretical work, we will investigate if it always discriminates correctly between ordered and chaotic motion in Hamiltonian systems, especially in dynamical systems with three degrees of freedom.

\section*{Acknowledgements}

I would like to express my warmest thanks to Dr. D.D. Carpintero for all the illuminating and creative discussions during this research. The author would also like to thank the two anonymous referees for the careful reading of the manuscript and for their very useful and aptly suggestions and comments which allowed us to improve significantly the quality and the clarity of the present article.

\end{document}